\documentclass[sigconf]{acmart}

\AtBeginDocument{%
  \providecommand\BibTeX{{%
    \normalfont B\kern-0.5em{\scshape i\kern-0.25em b}\kern-0.8em\TeX}}}

\copyrightyear{2025}
\acmYear{2025}
\setcopyright{cc}
\setcctype{by}
\acmConference[CHI '25]{CHI Conference on Human Factors in Computing Systems}{April 26-May 1, 2025}{Yokohama, Japan}
\acmBooktitle{CHI Conference on Human Factors in Computing Systems (CHI '25), April 26-May 1, 2025, Yokohama, Japan}\acmDOI{10.1145/3706598.3713454}
\acmISBN{979-8-4007-1394-1/25/04}

\usepackage{subcaption} 
\usepackage{url}
\usepackage{xcolor}
\usepackage{booktabs}
\usepackage{tabularx}
\usepackage{verbatim}
\makeatletter
\g@addto@macro{\UrlBreaks}{\UrlOrds}
\makeatother
\usepackage{multirow} %tabular cells spanning multiple rows
\usepackage{color} %colour management
\usepackage{enumitem} %better  enumerate, itemize and description

\usepackage{longtable}
\usepackage{xspace}
\usepackage{hyperref}

\newcommand{\notextrefrq}[1]{\hyperref[rq:#1]{\highlight{\textbf{\texttt{\textcolor{gray}{#1}}}}}}

\newcommand{\m}{\textit{M = }}

\newcommand{\pathfinder}{\textsc{PathFinder}\xspace}

\newcommand{\sd}{\textit{SD = }}

\usepackage{todonotes}

\usepackage{duckuments}
\usepackage{arydshln}

\usepackage{soul} % For highlighting
\definecolor{baseblue}{HTML}{2091f0} % Hex code for blue

\newcommand{\colorPone}{baseblue!47} % 90% transparent
\newcommand{\colorPtwo}{baseblue!16} % 75% transparent
\newcommand{\colorPthree}{baseblue!8} % 50% transparent
\newcommand{\colorPfour}{baseblue!8} % 25% transparent
\newcommand{\colorPfive}{baseblue!40} % 0% transparent

\newcommand{\Pone}{\sethlcolor{\colorPone}\hl{P1}\xspace}
\newcommand{\Ptwo}{\sethlcolor{\colorPtwo}\hl{P2}\xspace}
\newcommand{\Pthree}{\sethlcolor{\colorPthree}\hl{P3}\xspace}
\newcommand{\Pfour}{\sethlcolor{\colorPfour}\hl{P4}\xspace}
\newcommand{\Pfive}{\sethlcolor{\colorPfive}\hl{P5}\xspace}

\newcommand{\colorPoneStudy}{baseblue!8}    % P1: 0%
\newcommand{\colorPtwoStudy}{baseblue!8}    % P2: 0%
\newcommand{\colorPthreeStudy}{baseblue!80}% P3: 10%
\newcommand{\colorPfourStudy}{baseblue!8}   % P4: 0%
\newcommand{\colorPfiveStudy}{baseblue!16}  % P5: 1.5%
\newcommand{\colorPsixStudy}{baseblue!50}   % P6: 5%
\newcommand{\colorPsevenStudy}{baseblue!30} % P7: 3%
\newcommand{\colorPeightStudy}{baseblue!100}% P8: 14%
\newcommand{\colorPnineStudy}{baseblue!20}  % P9: 2%
\newcommand{\colorPtenStudy}{baseblue!100}  % P10: 10%
\newcommand{\colorPelevenStudy}{baseblue!60}% P11: 6%
\newcommand{\colorPtwelveStudy}{baseblue!8} % P12: 0%
\newcommand{\colorPthirteenStudy}{baseblue!80}% P13: 10%
\newcommand{\colorPfourteenStudy}{baseblue!30}% P14: 3.5%
\newcommand{\colorPfifteenStudy}{baseblue!90}% P15: 12%
\newcommand{\colorPsixteenStudy}{baseblue!10}% P16: 1%

\newcommand{\PoneStudy}{\sethlcolor{\colorPoneStudy}\hl{P1}\xspace}
\newcommand{\PtwoStudy}{\sethlcolor{\colorPtwoStudy}\hl{P2}\xspace}
\newcommand{\PthreeStudy}{\sethlcolor{\colorPthreeStudy}\hl{P3}\xspace}
\newcommand{\PfourStudy}{\sethlcolor{\colorPfourStudy}\hl{P4}\xspace}
\newcommand{\PfiveStudy}{\sethlcolor{\colorPfiveStudy}\hl{P5}\xspace}
\newcommand{\PsixStudy}{\sethlcolor{\colorPsixStudy}\hl{P6}\xspace}
\newcommand{\PsevenStudy}{\sethlcolor{\colorPsevenStudy}\hl{P7}\xspace}
\newcommand{\PeightStudy}{\sethlcolor{\colorPeightStudy}\hl{P8}\xspace}
\newcommand{\PnineStudy}{\sethlcolor{\colorPnineStudy}\hl{P9}\xspace}
\newcommand{\PtenStudy}{\sethlcolor{\colorPtenStudy}\hl{P10}\xspace}
\newcommand{\PelevenStudy}{\sethlcolor{\colorPelevenStudy}\hl{P11}\xspace}
\newcommand{\PtwelveStudy}{\sethlcolor{\colorPtwelveStudy}\hl{P12}\xspace}
\newcommand{\PthirteenStudy}{\sethlcolor{\colorPthirteenStudy}\hl{P13}\xspace}
\newcommand{\PfourteenStudy}{\sethlcolor{\colorPfourteenStudy}\hl{P14}\xspace}
\newcommand{\PfifteenStudy}{\sethlcolor{\colorPfifteenStudy}\hl{P15}\xspace}
\newcommand{\PsixteenStudy}{\sethlcolor{\colorPsixteenStudy}\hl{P16}\xspace}

\newcommand{\ignore}{}
\let\oldappendix\appendix
\renewcommand{\appendix}{\oldappendix\ignore}
\usepackage{etoolbox}
\pretocmd{\caption}{\ignorespaces\ignore}{}{}
\apptocmd{\caption}{\unskip}{}{}

\begin{document}

\title[Developing and Exploring a Multimodal Interface to Assist BVIPs to Exit HAVs]{Light My Way: Developing and Exploring a Multimodal Interface to Assist People With Visual Impairments to Exit Highly Automated Vehicles}

\author{Luca-Maxim Meinhardt}
\email{luca.meinhardt@uni-ulm.de}
\orcid{0000-0002-9524-4926}
\affiliation{%
  \institution{Institute of Media Informatics, Ulm University}
  \city{Ulm}
  \country{Germany}
}

\author{Lina Weilke}
\email{lina.weilke@uni-ulm.de}
\orcid{0009-0005-6689-2149}
\affiliation{%
  \institution{Institute of Media Informatics, Ulm University}
  \city{Ulm}
  \country{Germany}
}

\author{Maryam Elhaidary}
\email{maryam.elhaidary@uni-ulm.de}
\orcid{0009-0002-7658-0788}
\affiliation{%
  \institution{Institute of Media Informatics, Ulm University}
  \city{Ulm}
  \country{Germany}
}

\author{Julia von Abel}
\email{julia.von-abel@uni-ulm.de}
\orcid{0009-0003-4487-5458}
\affiliation{%
  \institution{Institute of Media Informatics}
  \city{Ulm}
  \country{Germany}
}

\author{Paul Fink}
\email{paul.fink@maine.edu}
\orcid{0000-0003-2915-1331}
\affiliation{%
  \institution{The University of Maine}
  \city{Maine}
  \country{US}
}

\author{Michael Rietzler}
\email{michael.rietzler@uni-ulm.de}
\orcid{0000-0003-2599-8308}
\affiliation{%
  \institution{Institute of Media Informatics}
  \city{Ulm}
  \country{Germany}
}

\author{Mark Colley}
\email{m.colley@ucl.ac.uk}
\orcid{0000-0001-5207-5029}
\affiliation{%
  \institution{Institute of Media Informatics}
  \city{Ulm}
  \country{Germany}
}
\affiliation{%
 \institution{UCL Interaction Centre}
 \city{London}
 \country{United Kingdom}
}

\author{Enrico Rukzio}
\email{enrico.rukzio@uni-ulm.de}
\orcid{0000-0002-4213-2226}
\affiliation{%
  \institution{Institute of Media Informatics, Ulm University}
  \city{Ulm}
  \country{Germany}
}

\renewcommand{\shortauthors}{Meinhardt et al.}

\begin{abstract}
The introduction of Highly Automated Vehicles (HAVs) has the potential to increase the independence of blind and visually impaired people (BVIPs). However, ensuring safety and situation awareness when exiting these vehicles in unfamiliar environments remains challenging. To address this, we conducted an interactive workshop with N=5 BVIPs to identify their information needs when exiting an HAV and evaluated three prior-developed low-fidelity prototypes. The insights from this workshop guided the development of \pathfinder, a multimodal interface combining visual, auditory, and tactile modalities tailored to BVIP's unique needs. In a three-factorial within-between-subject study with N=16 BVIPs, we evaluated \pathfinder against an auditory-only baseline in urban and rural scenarios. \pathfinder significantly reduced mental demand and maintained high perceived safety in both scenarios, while the auditory baseline led to lower perceived safety in the urban scenario compared to the rural one. Qualitative feedback further supported \pathfinder's effectiveness in providing spatial orientation during exiting.
% 149 words
\end{abstract}

\ccsdesc[500]{Hardware~Sensors and actuators}
\ccsdesc[500]{Human-centered computing~User studies}
\ccsdesc[300]{Human-centered computing~Laboratory experiments}
\ccsdesc[500]{Human-centered computing~Haptic devices}
\ccsdesc[300]{Human-centered computing~Sound-based input / output}
\ccsdesc[500]{Human-centered computing~Accessibility design and evaluation methods}
\ccsdesc[300]{Human-centered computing~Empirical studies in accessibility}
\ccsdesc[500]{Human-centered computing~Accessibility technologies}
\ccsdesc[500]{Human-centered computing~Accessibility systems and tools}

\keywords{people with visual impairments, multimodal interfaces, situation awareness, highly automated vehicles}

\begin{teaserfigure}
\centering
 \includegraphics[width=0.96\textwidth]{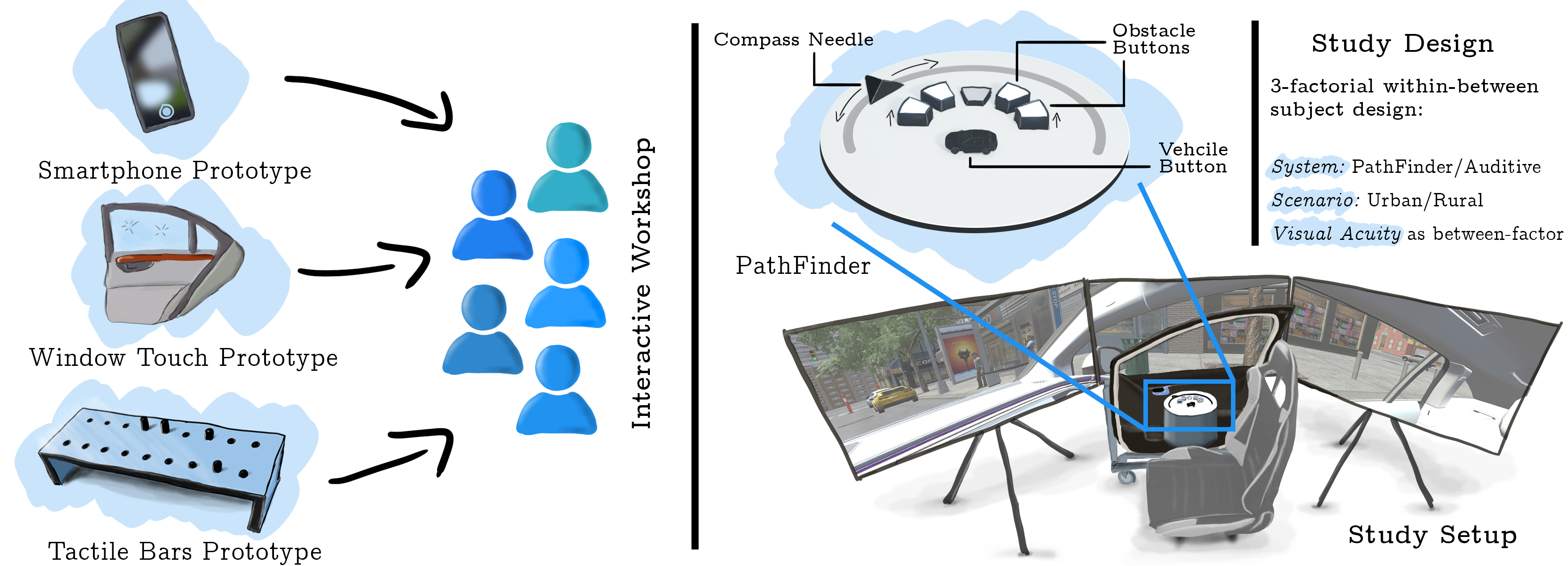}
  \caption{(Left) Interactive workshop (N=5) exploring the information needs of blind and visually impaired people when exiting future highly automated vehicles. Participants engaged with three initial low-fidelity prototypes: a smartphone, a window touch prototype, and tactile bars. (Right) Study setup featuring three monitors and a real car door with \pathfinder--a multimodal interface--attached to simulate a ride with an HAV. The top section explains \pathfinder's functionalities, including the compass needle, five extendable obstacle buttons, and the vehicle button. We used this setup to conduct a three-factorial within-between-subject study, using system and scenarios as our two within factors and participants' visual acuity as the between factor}
  \Description{This figure illustrates the three initial prototypes on the left side, each represented by a drawing. Arrows point from these prototypes toward five participant symbols, indicating that the prototypes were evaluated during an interactive workshop. On the right side, the PathFinder system is depicted, featuring a compass, five extendable obstacle buttons, and a vehicle button. The system is shown mounted between the car door and the seat. Behind this setup, three monitors display the surroundings of an HAV in the urban scenario}
  \label{fig:teaser}
\end{teaserfigure}

\maketitle

% Color Theme: #2091f0

\section{Introduction}
Over 270 million people worldwide live with vision impairments~\cite{WHO_blind, ackland2017world}, and this number is expected to rise as the population ages~\cite{Bourne.2020}. These impairments can limit daily activities such as driving~\cite{leeBlindness2024}, making independent mobility a significant challenge. Hence, the introduction of Highly Automated Vehicles (HAVs) in the near future has the potential to improve transportation for people who are blind or have visual impairments (BVIPs)~\cite{Brinkley.2017}. Studies indicate that sighted individuals barely expect increased independence using HAVs, but this expectation is significantly higher among BVIPs~\cite{Kacperski.2024}. 
Hence, by enabling independent and safe mobility among this demographic, HAVs represent a crucial step toward achieving greater equality in transportation~\cite{Bennett.2020}. However, ensuring safety and situation awareness when exiting these vehicles in unfamiliar environments remains a critical challenge as in today's manually driven vehicles, BVIPs often rely on drivers to drop them off at convenient locations that make it easier to navigate their surroundings~\cite{Brewer.2020}. With the introduction of HAVs, BVIPs may gain more independence~\cite{Kacperski.2024} but are likely to face situations alone without human assistance. This is where situation awareness will be particularly important as BVIP. This situation awareness involves ``\textit{the perception of the elements in the environment within a volume of time and space, the comprehension of their meaning, and the projection of their status in the near future}''\cite[p. 5]{endsleyTheorySituationAwareness1995}. In exciting situations, specific assistance—like detailed information about the vehicle's surroundings while parked—can benefit BVIPs~\cite{Brewer.2020, Fink.2023.Chauffeur}. 
However, situation awareness is not exclusive to BVIPs. Research shows that it also significantly enhances trust and perceived safety for sighted passengers in HAVs~\cite{You.2022, Hoff.2015, Woide.2022, Colley.2021, Pfleging.2021, faas2020longitudinal}.

% While BVIPs are optimistic about the benefits of HAVs~\cite{Fink.2021, Bennett.2020, Kacperski.2024}, they are concerned about whether HAVs will truly be designed to meet their needs~\cite{Bennett.2020, Brinkley.2020, Brinkley.2017}, particularly regarding their need for situational awareness of the vehicle's surroundings~\cite{Brinkley.2017, Brinkley.2020, Brinkley.2022.DesignTechniques, Brewer.2018.Understanding}.

Related research has already been conducted on parts of the trip via HAVs to enhance BVIPs' situation awareness, including locating ride-sharing vehicles with a smartphone application~\cite{Fink.2023.3} and enhancing situation awareness during rides by conveying traffic-relevant information~\cite{Fink.2023.2, fink_multisensory_2024, Meinhardt.2024, Md.Yusof.2020}. However, enhancing situation awareness and safety while exiting the HAV remains underexplored. Yet this part of the trip is crucial as it requires immediate awareness of potential hazards like moving cyclists~\cite{Brewer.2020} or obstacles that could cause trips or falls, posing significant safety risks. Unlike typical pedestrian navigation, where BVIPs rely on tools like canes or guide dogs, exiting an HAV involves rapidly adapting to a potentially unfamiliar and more hazardous environment. This situation requires new solutions to complement traditional navigation aids.

% This situational awareness involves ``\textit{the perception of the elements in the environment within a volume of time and space, the comprehension of their meaning, and the projection of their status in the near future}''\cite[p. 5]{endsleyTheorySituationAwareness1995}. 
% In these exciting situations, specific assistance—like detailed information about the vehicle's surroundings while parked—can be beneficial~\cite{Brewer.2020, Fink.2023.Chauffeur}. 
% Moreover, situational awareness is not exclusive to BVIPs. Research shows that it also significantly enhances trust and perceived safety for sighted passengers in HAVs~\cite{You.2022, Hoff.2015, Woide.2022, Colley.2021, Pfleging.2021, faas2020longitudinal}.

A notable attempt to address this research gap is the prototype ATLAS developed by \citet{Brinkley.2019}, which utilizes computer vision to articulate the surroundings upon arrival at the destination~\cite{Brinkley.2019}. Despite its advancements, such as increased trust towards the HAV, this solution is limited to auditory feedback only. However, incorporating additional modalities, such as tactile feedback~\cite{Yang.2022}, might be even more helpful by providing a multimodal approach~\cite{Fink.2023.2}. In fact, research suggests that integrating multiple modalities enriches the quality of information conveyed and significantly enhances situation awareness for BVIPs, offering advantages over single-modality feedback~\cite{Yatani.2012}. Specifically, the combination of voice-based and tactile feedback is particularly effective for navigation tasks~\cite{li_vision-based_2019}.

This paper explores a new interface designed to support BVIPs in such situations. Recognizing the advantages of multimodal interfaces in conveying information to BVIPs~\cite{Meinhardt.2024, Fink.2023.2}, we developed three initial prototypes (a smartphone, a window touch prototype, and tactile bars prototype) based on related work (e.g.~\cite{lee2022imageexplorer, guo_investigating_2018, holloway_accessible_2018, Meinhardt.2024}). Each prototype featured various modalities, including tactile, auditory, and visual cues, as well as different interaction strategies like pointing and sensing. 
This enabled us to conduct a focused evaluation of each modality and interaction strategy during an interactive workshop with N=5 BVIPs. In addition to evaluating the initial prototypes, the workshop explored the information needs of BVIPs when exiting a future HAV and possible methods to convey this information. 

The workshop results highlight the need for a multimodal approach to provide information about the vehicle’s surroundings. In response, we developed \pathfinder, a system designed to help BVIPs safely exit HAVs. By integrating visual, tactile, and auditory modalities into its design, \pathfinder adapts to BVIPs with different degrees of visual impairments. This approach ensures that \pathfinder effectively supports each passenger's individual needs in HAVs. 

We evaluated \pathfinder in a subsequent three-factorial, within-between-subject user study with N=16 BVIPs. This study compared \pathfinder to an auditory-only baseline, the current standard in accessible navigation technology, across two scenarios: a complex urban environment and a simpler rural setting. Quantitative results demonstrated that \pathfinder significantly reduced mental demand compared to the baseline. Additionally, the multimodal system consistently maintained high perceived safety in both scenarios, whereas the auditory baseline resulted in lower perceived safety in the urban scenario compared to the rural one. Additionally, qualitative feedback revealed a preference for multimodal information of conveyance of \pathfinder, which improved participant's spatial orientation.

\smallskip
\smallskip
\smallskip

\noindent\fcolorbox{blue}{blue!20}{\textbf{Contribution Statement}~\cite{Wobbrock.2016}}
\smallskip
\begin{itemize}[noitemsep, leftmargin=*]
    % \item \textbf{Artifact or System.} These prototypes were used as inspirational inputs for the interactive workshop, allowing our participants to explore various modalities and provide initial user feedback.
    
    \item \textbf{Empirical study that tells us about people.} We developed three low-fidelity prototypes with different interaction strategies and modalities to assist BVIPs in exiting HAVs, which were used as inspirational input for the interactive workshop with N=5 BVIPs. We found that participants preferred tactile cues as the basic modality to gain an overview of the surrounding HAVs, with auditory cues used for critical information, highlighting the need for multimodal accessible interfaces. 
    
    \item \textbf{Artifact or System.} Based on insights from the interactive workshop, we designed and developed \pathfinder, a multimodal interface including tactile, auditory, and visual modalities to assist BVIPs in exiting HAVs. This artifact demonstrates how the findings of the interactive workshop are applied to a concrete interface design that can be reproduced in future studies, as we provided all construction files as open source.

    % New knowledge is embedded in and manifested by artifacts and the supporting materials that  describe them.
    % Dies ermöglicht überhaupt eine Studie zu machen 
    % Es beantwortet *konkretiertes* Layout, aus wagen Ideen wird einen 
    % Nennen, dass es Open Source ist I'm Contribution Statements.
    
    \item \textbf{Empirical study that tells us about how people use a system.} In the following user study with N=16 BVIPs, we found that \pathfinder significantly reduced mental demand compared to an auditory-only baseline and maintained high perceived safety in both urban and rural scenarios. These results provide empirical evidence that multimodal interfaces can outperform unimodal systems in the HAV context, especially in complex environments, and highlight the need to tailor the interface to the user's visual acuity and the situation at hand.
\end{itemize}

\section{Related Work}
This research is grounded in current research on BVIPs and HAVs. We present navigation aids for BVIPs primarily designed to support pedestrians. Following this, we dive into the context of HAVs by describing current research on the needs of BVIPs within these vehicles.

\subsection{Navigation Aids for Visually Impaired People}
\citet{giudice2008blind} explored how technological aids assist with navigation for people with visual impairments, identifying four key considerations: (1) The conveyance of visual information into auditory or tactile modalities should be defined clearly, accommodating the cognitive demands and learning curve of users. (2) The presented information should be minimized to the essentials. (3) Given that each system has unique advantages and disadvantages depending on the context, combining various aids might be necessary for effective navigation across different scenarios. (4) Devices should be designed to be non-intrusive and aesthetically pleasing.
Building on these guidelines, several navigation aids have been developed and assessed to support BVIPs. For this, \citet{Ducasse.2018} reviewed various dynamic tactile maps for BVIPs, classifying them into \textit{Digital Interactive Maps} displayed on flat surfaces like screens and \textit{Hybrid Interactive Maps} that incorporate both digital and physical elements. These dynamic tactile maps have demonstrated higher performance compared to touchscreen and swell paper maps (a type of tactile paper that raises printed images or text) regarding map reading speed and the ability to create a mental map of the route~\cite{Zeng.2015}. Further research by \citet{Holloway.2019} demonstrated that tactile maps \textit{[...] support orientation and mobility through identification of landmarks, route planning and creation of a mental map [...]}~\cite[p.184]{Holloway.2019}.

In general, multimodal approaches to conveying information seem to outperform those that rely on a single modality, as supported in \textit{multiple-resource theory} by \citet{wickens_compatibility_1983}, which states that distributing information across modalities such as auditory and visual reduces competition for attention and processing resources, leading to better task performance, and reduced mental demands as task difficulty increases~\cite{oviatt_when_2004}. Neuroimaging studies further support this, showing that the occipital cortex in blind individuals represents spatial information similarly across different sensory inputs~\cite{amedi_occipital_2005}, facilitating sensory-independent spatial representations~\cite{loomis_spatial_2002}. Therefore, multimodal interfaces leverage cognitive advantages and neural adaptability, potentially leading to more effective navigation aids for BVIPs. This argument is in line with finding from \citet{kuriakose_tools_2022}, who reviewed multiple tools and technologies that support BVIPs in their navigation task, recommending that ``\textit{if there is an option for multiple feedback modalities, the user will get the flexibility to choose one based on a situation or environment}''~\cite[p. 12]{kuriakose_tools_2022}. This aligns with \citet{Yatani.2012}, who found that handheld tactile maps combining tactile feedback with audio instructions offer superior spatial orientation compared to audio-only feedback. Additionally, the study revealed differences in the effectiveness of verbal audio vs. auditory icons, aligning with the findings of \citet{Glatz.2018}, who found auditory icons to be more effective for conveying contextual information, while verbal audio was better for urgent requests. Further, by comparing the effectiveness of auditory, visual, and combined audio-visual feedback, the combination of audio and visual feedback improved participants' situation awareness more than visual feedback alone~\cite{Nadri.2021}. Additionally, multimodal maps with tactile elements, augmented by audio feedback when touched, enhanced navigation skill improvement for BVIPs. Participants especially valued the combination of audio and tactile cues, highlighting the importance of designing such tools in line with users’ preferences and needs \cite{albouys-perrois_towards_2018}.
% We have a few papers on vibro-audio maps as well that might help here, with strong support for equivalent cognitive map development between blind and sighted users when BLVPs are given matched information using VAMs (Palani et al., 2022) that have real-world relevance for solving navigation problems for BLVPs in the transportation domain with HAVs (Fink et al., 2024). 

% Palani, H. P., Fink, P. D. S., & Giudice, N. A. (2022). Comparing Map Learning between Touchscreen-Based Visual and Haptic Displays: A Behavioral Evaluation with Blind and Sighted Users. Multimodal Technologies and Interaction, 6(1), Article 1. https://doi.org/10.3390/mti6010001

% Fink, P. D. S., Milne, H., Caccese, A., Alsamsam, M., Loranger, J., Colley, M., & Giudice, N. A. (2024). Accessible Maps for the Future of Inclusive Ridesharing. Proceedings of the 16th International Conference on Automotive User Interfaces and Interactive Vehicular Applications, 106–115. https://doi.org/10.1145/3640792.3675736

Given the advantages of multimodal systems for navigation tasks and following the guidelines of \citet{giudice2008blind}, this work investigates the potential of multimodal interfaces for BVIPs in the automotive domain. The following sections will explore the specific needs of BVIPs inside HAVs, providing a foundation for developing new systems tailored to their requirements.

% It's good you have the Oviatt citation. I recommend also mentioning that multimodal information is especially relevant for spatial information (such as the exiting task used here), given that spatial information is represented similarly in the brain across sensory inputs (Amedi et al., 2005), contributing to sensory-independent representations (Loomis et al., 2002) that support equivalent behaviors for sighted and blind people (Wolbers et al. 2011; Loomis et al., 2013). 
% Amedi, A., Merabet, L. B., Bermpohl, F., & Pascual-Leone, A. (2005). The Occipital Cortex in the Blind: Lessons About Plasticity and Vision. Current Directions in Psychological Science, 14(6), 306–311. https://doi.org/10.1111/j.0963-7214.2005.00387.x
% Loomis, J. M., Lippa, Y., Klatzky, R. L., & Golledge, R. G. (2002). Spatial updating of locations specified by 3-D sound and spatial language. Journal of Experimental Psychology: Learning, Memory, and Cognition, 28(2), 335–345. https://doi.org/10.1037/0278-7393.28.2.335
% Wolbers, T., Klatzky, R. L., Loomis, J. M., Wutte, M. G., & Giudice, N. A. (2011). Modality-Independent Coding of Spatial Layout in the Human Brain. Current Biology, 21(11), 984–989. htps://doi.org/10.1016/j.cub.2011.04.038
% Loomis, J. M., Klatzky, R. L., & Giudice, N. A. (2013). Representing 3D space in working memory: Spatial images from vision, hearing, touch, and language. In Multisensory imagery (pp. 131–155). Springer Science + Business Media. https://doi.org/10.1007/978-1-4614-5879-1_8

\subsection{Needs and Opinions of Visually Impaired People in the Context of Highly Automated Vehicles}
While the aforementioned studies primarily focused on pedestrian navigation, the introduction of HAVs presents new opportunities and challenges for BVIPs. Most BVIPs are enthusiastic about the autonomy HAVs promise, potentially granting access to previously challenging locations~\cite{Kacperski.2024}. Despite this, initial qualitative research showed that BVIPs raised concerns regarding whether HAVs will be truly designed to meet BVIPs needs~\cite{Bennett.2020, Brinkley.2020, Brinkley.2017}. To envision these needs, current rideshare services have been used as a proxy for future HAV scenarios~\cite{Brewer.2020, Fink.2023.Chauffeur} and workshops have been conducted to identify passengers' needs and imagine accessible interfaces for HAVs~\cite{Brewer.2018.Understanding}. Results have demonstrated the need for non-visual support for BVIPs throughout the complete transportation trip via HAVs, from locating an HAV~\cite{fink2023accessible} to conveying traffic information, such as the reason for the HAV stopping during the ride~\cite{Meinhardt.2024, Fink.2023.2}. The following sections review the small but emerging field of literature on nonvisual interfaces across the complete trip via HAVs.

\subsection{Non-visual Interface Development Across the Entire Journey}
Initial research has sought to design and test accessible interfaces in HAVs. For instance, researchers have evaluated mid-air haptics and tactile interfaces to enhance situation awareness during the ride~\cite{Fink.2023, Meinhardt.2024}. 
% Additionally, gestural interactions have been explored for in-vehicle control by BVIPs~\cite{Fink.2023.Chauffeur}. 
While these in-vehicle studies have shown promising results, they have primarily focused on the on-road part of the trip. Only a few studies have investigated other parts of the trip, such as pre-journey mapping~\cite{Fink_AUto_UI} and vehicle localization~\cite{Fink.2023.3, RANJBAR2022100630}. While the ATLAS system by \citet{Brinkley.2019} explored supporting BVIPs in gaining situation awareness when exiting the vehicle, it solely utilized the auditory modality, missing the benefits of multimodal interfaces that can support BVIPs inside HAVs~\cite{Fink.2023.2, Meinhardt.2024}. \\
To address this research gap, this work will examine BVIPs' information needs for exiting HAVs and investigate the potential of a multimodal interface to assist them during this phase. To achieve this, we hosted an interactive workshop with BVIPs and created three initial low-fidality prototypes, which will be described in the following section.
% To achieve this, we created three low-fidelity prototypes and presented them to a group of N=5 BVIPs as a starting point for discussion during a workshop. This interactive workshop aimed to explore the specific information needs and the most effective ways to convey this information. The following sections will describe these low-fidelity prototypes and their integration into the interactive workshop.

\section{Initial Low-Fidelity Prototypes}\label{sec:initial_prototypes}

\begin{figure*}[ht!]
\centering
\small
    \begin{subfigure}[c]{0.325\linewidth}
        \includegraphics[width=\linewidth]{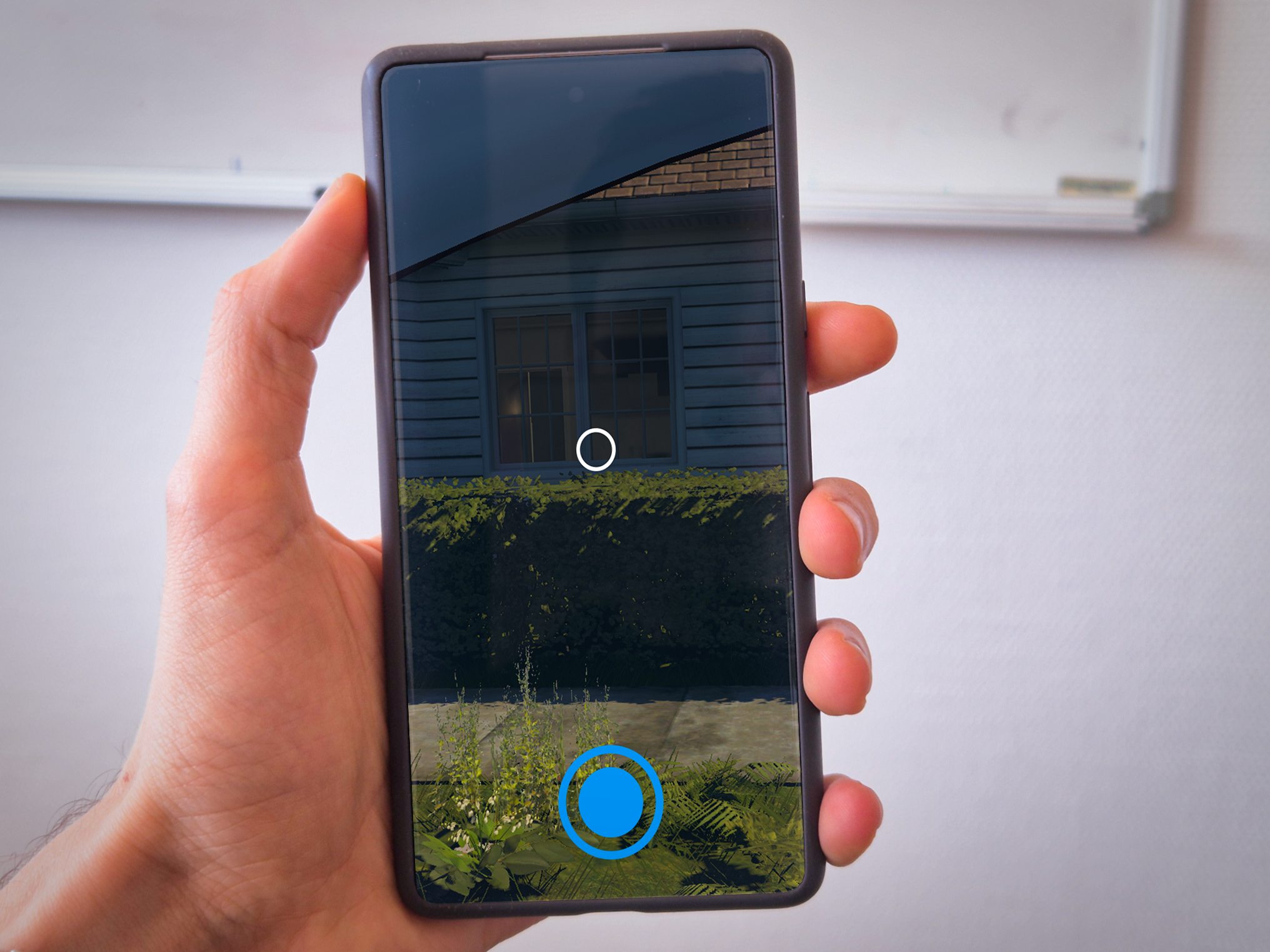}
        \caption{Smartphone Prototype}\label{fig:smartphone_prototype}
        \Description{This figure depicts a hand holding a smartphone. On the smartphone screen, a virtual environment is displayed, simulating a real-time view of the surroundings of the HAV as if someone had taken a photo from inside the vehicle}
    \end{subfigure}
    \hspace{0.1em}
    \begin{subfigure}[c]{0.325\linewidth}
        \includegraphics[width=\linewidth]{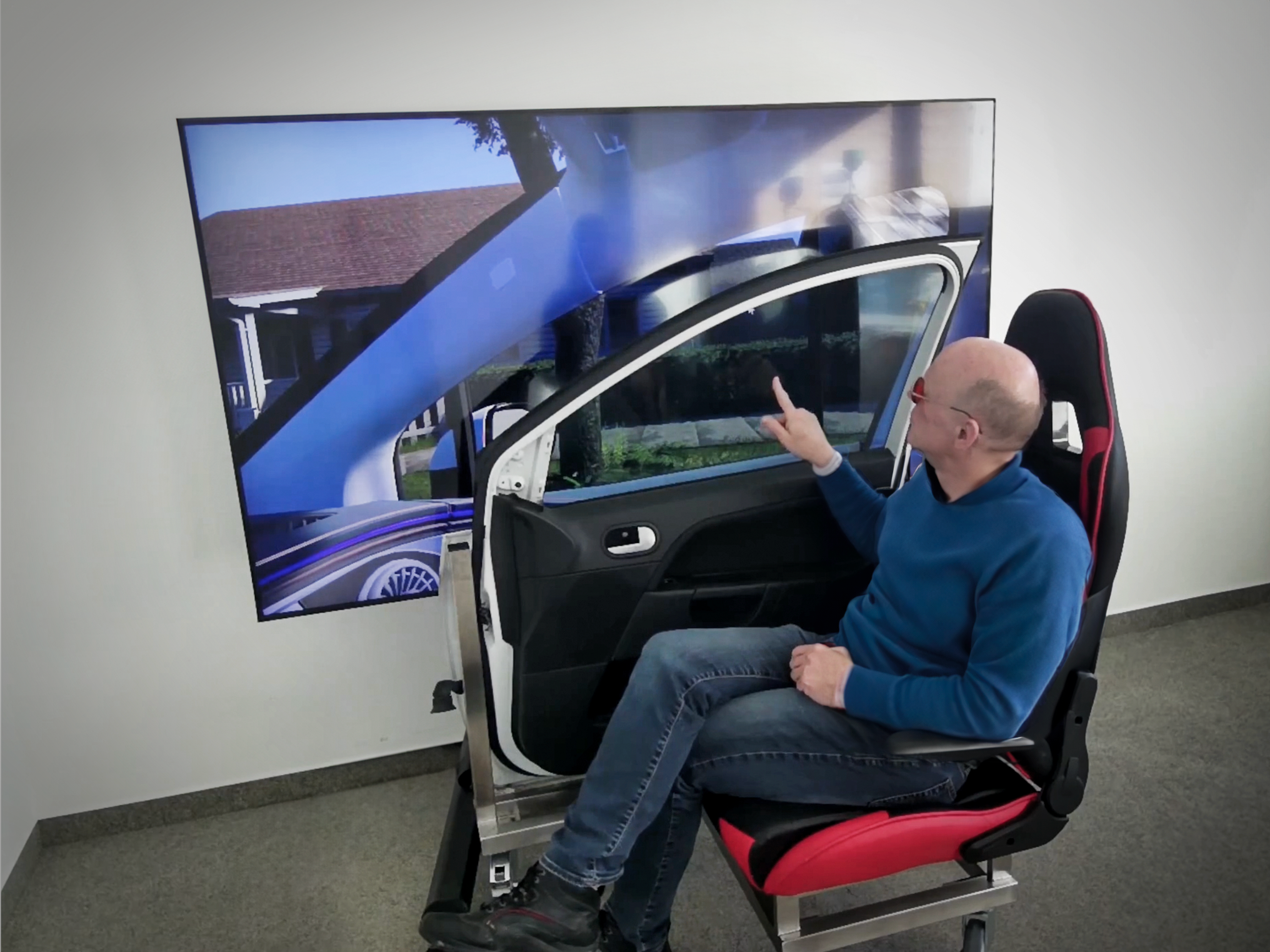}
        \caption{Window Touch Prototype}\label{fig:car_window_prototype}
        \Description{This figure shows a workshop participant seated in a car seat, positioned in front of a car door. Behind the window, a display simulates the surroundings of an HAV}
    \end{subfigure}
    \hspace{0.1em}
    \begin{subfigure}[c]{0.325\linewidth}
        \includegraphics[width=\linewidth]{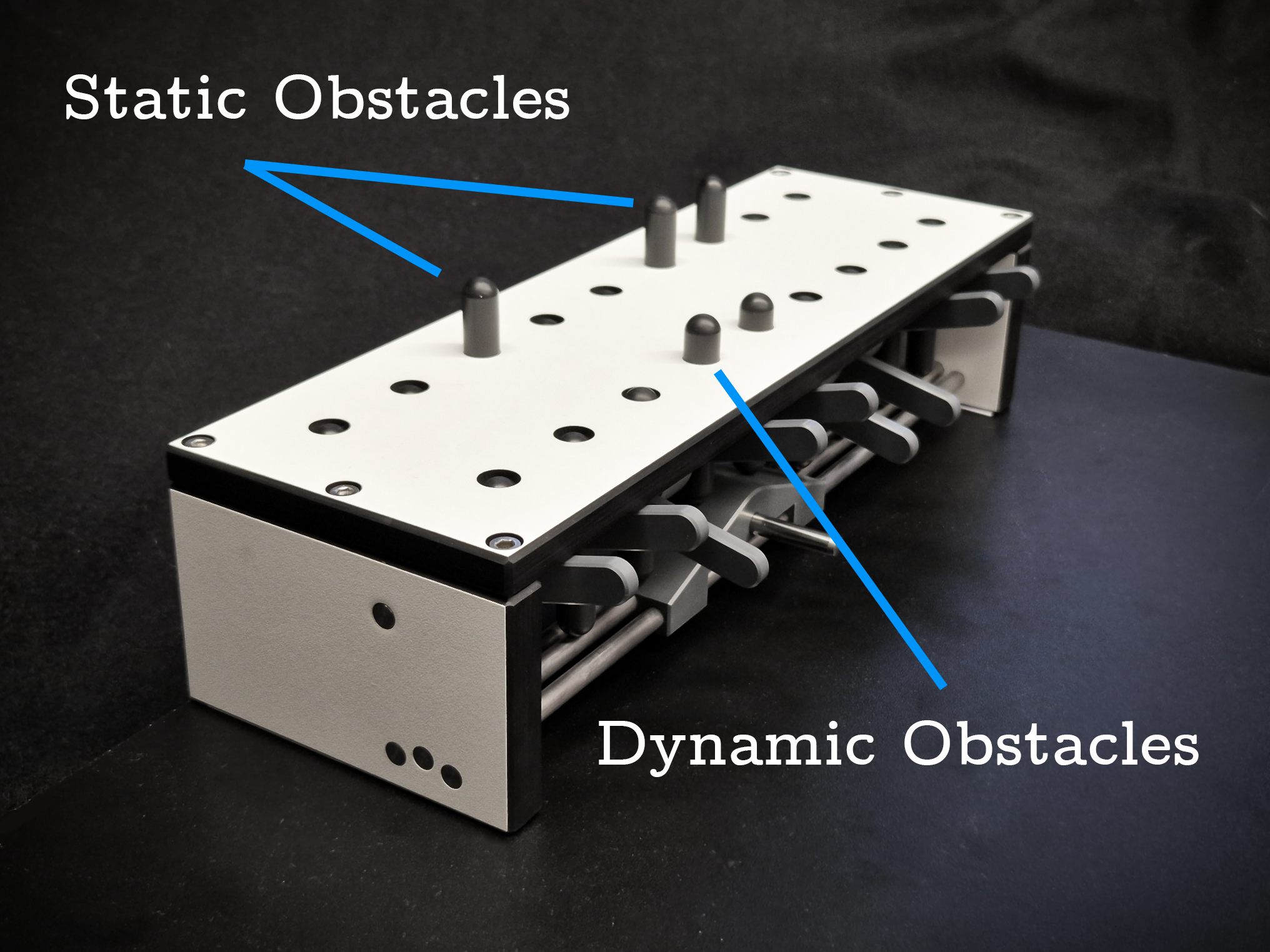}
        \caption{Tactile Bars Prototype}\label{fig:haptic_prototype}
        \Description{This figure shows the tactile bar prototype, which is a white box featuring two rows of black bars or pins. These bars can be moved upward using handles located on the sides of the prototype}
    \end{subfigure}
    \hspace{0.1em}

   \caption{Initial low-fidelity prototypes that where used during the interactive workshop}~\label{fig:initial_prototypes}
   \Description{Three pictures of the initial low-fidelity prototypes}
\end{figure*}

Prior to hosting the interactive workshop with N=5 BVIPs, we developed three initial low-fidelity prototypes (smartphone, window touch, and tactile bars) based on existing research~\cite{lee2022imageexplorer, guo_investigating_2018, holloway_accessible_2018, Meinhardt.2024} and state-of-the-art smartphone applications for BVIPs~\cite{guo_investigating_2018, Fink.2023.2, be_my_eyes, seeingaiSeeingTalking}. This section will detail each prototype and explain the related work from which they were derived. 
The three prototypes were designed to serve as concrete examples to inspire and facilitate discussion during the workshop, providing participants with tangible prototypes to interact with rather than relying solely on conceptual discussions about potential interaction strategies and modalities. Hence, each prototype employs different modalities and interaction strategies. This approach enabled us to evaluate each modality and strategy independently in a focused manner during the subsequent workshop. Below, we describe each prototype in detail, along with the rationale for their design choices. Additionally, we provided the construction files for each initial prototype for reconstruction in a git repository (see section \nameref{sec:openscience}).

\paragraph{Scenario Design for the Prototypes}
For the prototypes, we designed a simulated suburban scene using Unity~\cite{unitygameengine} version 2023.2.1f1 and the Suburb Neighborhood House Pack asset~\cite{suburbasset}. This scene, which was used in both the smartphone prototype and the window touch prototype, depicted an HAV parked on the side of the road, with a pedestrian/cyclist path next to it and a house (the final destination) behind the path. The scene included static obstacles such as a tree and a street sign near the HAV's door, as well as dynamic obstacles like pedestrians and cyclists moving in front of the door.

\subsection{Smartphone Prototype}
Smartphones are prevalent among BVIPs, especially among young BVIPs (19-34); 76\% of them own a smartphone~\cite{abraham_smartphone_2022}. 
Building on this familiarity, we developed a smartphone prototype inspired by previous work, such as the object detection application by \citet{zhong_real_2013} and more recent smartphone-based navigation aids in the automotive context for BVIPs~\cite{Fink.2023.3}. 
Research has indicated that BVIPs prefer to move the smartphone to scan their surroundings when exploring their environment, compared to other interaction strategies via smartphone~\cite{guo_investigating_2018}. Therefore, 
our smartphone prototype (\autoref{fig:smartphone_prototype}) allows users to scan their surroundings using the smartphone's camera, triggering auditory descriptions of static objects and dynamic obstacles within the simulated suburban scene.
A button on the screen allows participants to cast a ray within the Unity environment, identifying objects like "Tree" via verbal auditory feedback. Vibration feedback confirms successful button activation, enhancing interaction~\cite{palani_design_2020}. Dynamic obstacles, such as approaching cyclists, are automatically announced in real-time to ensure immediate awareness.\\
The interaction strategy combines verbal and auditory modalities, supported by visual feedback through the smartphone's scanning and pointing mechanism.

\subsection{Window Touch Prototype}\label{sec:carwindowprototype}
Our window touch prototype ( \autoref{fig:car_window_prototype}) draws inspiration from Ford's "Feel the View" system~\cite{feelTheView}, which allows users to receive tactile feedback about the outlines of the environment on the vehicle's side window. 
Our prototype extends this concept, but instead of using vibration, we employed verbal auditory feedback of the obstacles when the participants touch the window, as based on touch-exploration of images~\cite{lee2022imageexplorer, seeingaiSeeingTalking}.

To demonstrate this prototype, we created a setup with a car door and a 75" monitor displaying the same suburban scene used in the smartphone prototype. This setup was designed as a Wizard-of-Oz prototype, where one of the workshop moderators manually triggered the corresponding verbal output upon participants' pointing. For example, if a participant pointed toward a tree, the verbal sound "Tree" was played. Like the smartphone prototype, information about dynamic obstacles, including their direction, was communicated automatically upon the HAV reaching its destination.\\
This prototype focuses on a verbal, auditory modality, enhanced by touch-based interaction, allowing participants to receive detailed information about their surroundings directly through the car window.

\subsection{Tactile Bars Prototype}
Related work showed that tactile cues can help gain situation awareness for BVIPs inside HAVs~\cite{Meinhardt.2024, Fink.2023, Fink.2023.2, Md.Yusof.2020, Chiossi.2022, Sonoda.2017}. Hence, we designed a tactile bar prototype (see \autoref{fig:haptic_prototype}) to convey potential obstacles when exiting the vehicle. Unlike the smartphone and window touch prototypes, this one does not rely on a visual Unity scene; instead, it solely uses tactile feedback to convey information about obstacles. 

The tactile bars prototype features two rows of nine movable bars. The first row (from the perspective of the participants) represents static obstacles, such as trees or street signs, while the second row represents dynamic obstacles, like cyclists or pedestrians. We rounded the edges of the bars to ensure a smooth surface to avoid discomfort when touching them, as recommended by \citet{holloway_accessible_2018}. The rationale for having two distinct rows is to separate the types of obstacles, assuming to make it easier for participants to understand the environment. Static obstacles are presented as constant and unchanging, while dynamic obstacles are represented with motion, created by manipulating the bars in the second row to simulate movement.

In the first row, each bar is controlled by one of nine levers, operated by one of the workshop moderators, allowing individual movement up and down. A slider at the prototype's bottom manipulates the dynamic obstacle bars. Moving this slider creates a wave-like effect on the bars, creating a tactile illusion of motion. The decision to convey motion via bars that move up and down was inspired by \citet{holloway_animations_2022}, who noted that using height differences is a perceivable method to convey tactile motion for BVIPs. \\
The prototype's design prioritizes simplicity and tactile feedback, offering BVIPs to gather information about their surroundings without relying on visual or auditory cues.

\section{Interactive Workshop}\label{workshop}
In this section, we describe the interactive workshop we conducted with N=5 BVIPs (three female, two male and no non-binary) aged between 44 to 67 (\m{57.80}, \m{8.61}). The female participants reported being completely blind, with one having light perception. The male participants had impaired vision, with visual acuity between 3-5\% (see \autoref{app:demogr} for more details). 

The workshop was conducted to identify BVIPs' specific information needs to improve their situation awareness and assist in safe exit of HAVs. Further, we explored preferred interaction strategies to convey the necessary information for these tasks. To provide prior inspiration and a starting point for discussion, the three initial low-fidelity prototypes (see \autoref{sec:initial_prototypes}) were presented to the participants in individual sessions during the workshop. \\
The following section details the interactive workshop procedure and the implications of the results, which eventually led to the final design of \pathfinder. By including the participants from the beginning of the design phase, we adopted the Participatory Design approach from \citet{ParticipatoryDesign.1993}.

\subsection{Procedure}\label{ch:Workshop_Procedure}
The workshop, moderated by five of the authors, was scheduled to last three hours and divided into four phases. The detailed agenda for the session is outlined in \autoref{tab:agenda}. Before the start, we ensured that all participants had consented to share their data, which allowed us to proceed with audio and video recordings during the study.

During the first phase, we started with brief introductions, where each participant shared their visual abilities. This was followed by a concise overview of the capabilities of HAVs, which can reach their destination without any intervention, as defined by SAE levels 4 and 5~\cite{SAElevel}.

Transitioning into the workshop's core discussion in the second phase, we opened the floor to conversations about the participants' personal experiences with exiting traditional vehicles, such as taxis. Following this, we asked participants about their specific information needs to gain situation awareness of the surrounding environment. Next, we invited participants to generate ideas on how to convey these information needs to them.

\begin{table*}[ht!]
 \small
     \caption{Scheduled agenda for the three hours interactive workshop}
     \Description{Scheduled agenda for the three hours interactive workshop divided into four phases, staring with the introduction of the participants, the group discussions, and the individual interactive sessions with the initial prototypes}
   \label{tab:agenda}
  \begin{tabular}{ccl}
    phase & scheduled duration & agenda \\
    \midrule
    1 & 15 min & Introduction of the Participants and Moderators (4 authors). Overview of HAVs capabilities \\    
    2 & 45 min & Open Group Discussion about Information Needs and Information Conveyance to Exit the Vehicle \\  
    3 & 75 min & Individual Interactive Prototype Sessions (15 min per Participant and Prototype)  \\  
    4 & 45 min & Open Group Discussion about the Prototype Interaction \\
  \end{tabular}
\end{table*}

After collecting their unbiased ideas, we moved to the third phase: the interactive prototype sessions. Here, we presented the three initial prototypes to each participant individually in a counterbalanced order. This approach ensured that the participants' initial ideas remained independent of our prototypes, thus avoiding bias in their creativity.
During each individual interactive prototype session, we prompted the participants with an initial story to envision themselves in an HAV, traveling alone to a friend's house for the first time, simulating their unfamiliarity with the area. While interacting with the prototypes, we asked the participants to perform the Thinking-Aloud method~\cite{jaspers_think_2004}. Hence, we asked participants to explain their thoughts about each part of the prototype, describing what they thought it represented and what aspects of the information conveyance they liked or found difficult, including their reasons.
Given that the prototypes were set up in three separate rooms, we arranged for three participants to interact with the prototypes simultaneously while the remaining two participants waited. Each prototype was operated by one of the moderators. During the sessions, we briefly explained the prototype's interaction strategies and let the participants interact with the prototype while thinking aloud. 

After the prototype sessions, we gathered the participants again in a group discussion to ask them about their positive and negative experiences with the prototypes and their interaction strategies (phase four).
By first collecting individual impressions during the prototype sessions, we ensured that the feedback remained honest and unbiased from the other participants, facilitating diverse viewpoints from the group.
The participants were compensated with 30 \anon{Euros} for their time during the three-hour workshop.

\subsection{Analysis}
Four authors conducted a reflexive, inductive thematic analysis, following the approach of \citet{Clarke.2006, Clarke.2020}. We analyzed audio and video recordings from the workshop, focusing on both group discussions and think-aloud sessions of the prototypes. The codes generated from this analysis were organized on a digital whiteboard, sorted by feedback for each prototype and the group discussions. We then grouped these codes into thematic clusters before moving to the third phase of thematic analysis: searching for themes.
This was done in a group meeting among the authors. In cases of disagreement, we engaged in discussions to resolve any discrepancies. In total, we generated 396 codes from the interactive workshop, which were clustered into 8 subclusters and three main themes.

\subsection{Results and Implications}
Our findings are divided into three main sections based on the identified themes: (1) current situations to exit a vehicle, (2) the information needs of participants when exiting HAVs, and (3) methods for effectively conveying this information. The first section mainly derives from the open group discussion about information needs. The second section derives from the individual interactive prototype sessions and the subsequent group discussion (see \autoref{tab:agenda}). However, before we dive into these two key areas, we first provide an overview of the participants’ current experiences when exiting a vehicle. \\
To correlate participants' visual acuity with our findings, we used blue highlighting with different levels of transparency. The transparency level reflects each participant's visual acuity: participants with lower visual acuity, like \Pthree\ and \Pfour\ (0\%), had more transparent highlighting, while those with higher visual acuity, like \Ptwo\ (1\%), \Pfive\ (3\%), and \Pone\ (5\%), had darker, less transparent highlighting. For more detailed demographic information, please refer to \autoref{app:demogr}.

\subsubsection{Current Situation to Exit a Vehicle}
All participants consistently mentioned their reliance on the assistance of others, such as taxi drivers, when exiting the vehicle. For example, \Ptwo shared, ``\textit{I rely on the taxi driver to guide me until I am familiar with my surroundings again}''. \Pfour agreed, adding that she also asks if it is safe to open the car door before exiting. During the exit, she holds the cane with her right hand while using her left hand for support.
The participants generally relied on more assistance to exit the vehicle in an unfamiliar environment, as mentioned by \Pone.
Further, \Pfive emphasized the value of communicating his visual impairment in an unfamiliar vehicle. He explained that sharing information about his condition enhances his perceived safety and ensures that others are mindful of his needs. Likewise, both \Pfive and \Pthree mentioned that being in the company of acquaintances increases their perceived safety, as these people are already familiar with their needs. Many of the insights align with \citet{Brewer.2020} whose participants stated that ``\textit{they asked drivers to drop them off at convenient locations that made it easier to find doors}''~\cite[p. 3]{Brewer.2020}.

Recognizing the current need for assistance in exiting the vehicle to gain situation awareness of the environment is essential. The potential increasing independence of BVIPs with the introduction of HAVs~\cite{Kacperski.2024} highlights the need for interfaces that support BVIPs in exiting future HAVs. By exploring and understanding the specific information needs when exiting HAVs, we can contribute to the design of future HAVs that promote not only accessibility but also independence and perceived safety.

\subsubsection{Information Needs When Exiting HAVs} \label{sec:informationNeeds}
\Pfive summarized that when exiting a potential HAV ``\textit{it is important to find out immediately what [obstacle] it is, and then I can decide whether it is relevant for me}''. Echoing this statement, \Pone and \Pthree acknowledged that while technology can aid them in gaining situation awareness, they still feel responsible for their actions and strive to maintain their sense of control, as already suggested by \citet{Brewer.2018.Understanding}. However, they highlighted the critical need for direct communication in potentially dangerous situations, such as cyclists passing in front of the HAV before exiting. 
Once their situation awareness needs are met, participants noted no further information requirements after leaving the vehicle. \Pone clarified, ``\textit{As soon as I leave the car, that's my concern, but I know which way I'm going}.'' Reflecting a similar sentiment, \Pfour and \Ptwo mentioned their preferred reliance on traditional mobility aids, such as canes or guide dogs, immediately after exiting the HAV.
Diving into the concrete information needs, we categorized the participants' needs into five categories: (1) static obstacles, (2) dynamic obstacles, (3) the condition of the ground, (4) information needs about the final destination, and (5) the spatial orientation.

\textbf{Static Obstacles.} 
Our workshop participants mentioned multiple static objects they would need to be informed about when exiting HAVs, such as trees in front of the pedestrian path, road signs, garbage cans on the road, road bollards, or parking vehicles. Further, for \Pthree and \Pone, the information about a safe pedestrian path is crucial. Opinions on the need for information about the distances to these obstacles were mixed. \Pone, who has relatively high visual acuity among the group, expressed that he could independently estimate these distances and did not require explicit information. Conversely, \Pfive, despite having similar visual acuity, preferred to have distances explicitly communicated, aligning with the other participants' preferences.

\textbf{Dynamic Obstacles.} 
All participants agreed that information about dynamic obstacles, such as cyclists passing in front of the vehicle, is crucial. \Pone specifically noted, ``\textit{Very fast cyclists are frightening; they don't take any care of me}''. \Pthree added that knowing the direction of these dynamic obstacles is essential. She would also appreciate information about when an obstacle has passed. Further, \Pfour emphasized the importance of receiving updates about dynamic obstacles just before exiting the vehicle, as this timing is most critical for her situation awareness.

\textbf{Terrain Perturbations.}
In addition to static and dynamic obstacles, participants highlighted the importance of understanding the ground conditions around the vehicle. In particular, they noted that awareness of potentially dangerous surfaces, such as slippery ice or wet grass, is critical given the increased risk of injury from such conditions. However, \Pfive mentioned that while this information might be important for people with total blindness, he would not require this kind of information.

\textbf{Information Needs About the Final Destination.}
All participants expressed the need for detailed information about their final walking destination. \Pthree specified the importance of knowing the approximate distance and direction to the final destination. \Pone expanded on this, highlighting its particular significance in unfamiliar environments. He stressed that understanding which side of the vehicle to exit from and the route to the final destination are his highest priorities among all information needs.

\textbf{Spatial Orientation.}
All participants emphasized acquiring spatial orientation before exiting the vehicle to improve their situation awareness. In this context, \Pthree pointed out that ``\textit{if you become blind later in life, your spatial perception differs from someone who has been blind since birth}''. This aligns with the deficiency model by  \citet{vonSenden.1960} arguing that visual experience is critical for accurate spatial orientation. Accordingly, the lack of visual experience slows down and reduces the accuracy of situation awareness for BVIPs, leading to less spatial orientation compared to sighted individuals~\cite{vonSenden.1960}. More recent studies however indicate that BVIPs are able to gain the same spatial orientation as sighted people when enough information is provided~\cite{loomis_spatial_2002, lacey_representing_2013}.
Thus, to increase the amount of information to gain spatial orientation, \Pfour, \Pthree, and \Pfive all agreed on the importance of using the ego vehicle as a reference point to contextualize other objects in the environment. Additionally, \Pthree suggested that information should be organized in a structured manner (e.g., arranged in a circle) to support her spatial orientation.

\subsubsection{Information Conveyance}\label{sec:information_conveance}
The insights on how to convey the information needs discussed previously were mainly derived from individual and group feedback on the initial low-fidelity prototypes (see \autoref{fig:initial_prototypes}).

\textbf{Active and Passive Interaction.}
In general, participants preferred receiving crucial information passively rather than seeking it out actively, as with the smartphone prototype. For instance, \Pfive noted his discomfort with actively scanning the surroundings. He also pointed out that relying solely on auditory feedback would be insufficient in scenarios where other passengers are talking within the HAV, thus expressing a preference for the tactile bars prototype in this situation.

Feedback on the auditory and tactile modalities varied among participants. \Pfive found the tactile bars prototype helpful for gaining a broad initial overview of the environment, though he noted it was insufficient for detailed information. He explained, ``\textit{I often drive with noisy children. Tactile output would let me [actively] sort out important details like necessary precautions by myself},'' supporting the findings of \citet{DiCampliSanVito.2019} that tactile feedback is less distracting and bothersome than other interfaces. \Ptwo and \Pthree suggested integrating voice output similar to the window touch and smartphone prototype to enhance the tactile bars prototype. This suggestion echoes the ISANA system from \citet{li_vision-based_2019} to enhance BVIPs' navigation tasks. Nevertheless, there was a consensus among all participants that critical information, such as cyclists should primarily be conveyed passively through voice. Additionally, \Pfour preferred that voice output be as concise as possible. This requirement underscores the importance of delivering clear and succinct information to avoid overwhelming the passengers with excessive details.
Further, many found the smartphone prototype cumbersome and inconvenient. For example, \Pfour mentioned, ``\textit{I found that a bit stupid with the smartphone; I don't have enough hands for it when I get out of the car.}'' Most participants (4 of 5) shared this sentiment, indicating discomfort with not having their hands free.

\textbf{Completeness of Information.}
All participants emphasized the importance of being informed when all relevant information were conveyed to them. This requirement was well met by the tactile prototype, as \Ptwo and \Pthree could physically sense when they had explored all available information with their hands. However, the smartphone prototype presented some challenges; for example, \Ptwo criticized the absence of a physical boundary or frame to guide the smartphone's movement. Similarly, \Pthree expressed difficulties in effectively scanning the environment, remarking, ``\textit{I have to scan the environment, but I'm imperfect}.'' These issues were also reflected in using the window touch prototype, where \Pfour was uncertain about whether she had touched all relevant objects on the window.
To overcome these challenges, \Pfive suggested implementing a standardized output of information to ensure passengers are consistently aware when all relevant data has been communicated.

\textbf{Variations in Different Visual Acuities}
Participants' responses to the initial prototypes varied significantly based on their visual acuity. \Pone, who retains 5\% visual acuity, expressed discomfort with being overwhelmed by excessive information. In contrast, \Ptwo (1\% acuity) advocated for providing more information rather than less, allowing passengers the autonomy to determine which details are relevant to their needs. Furthermore, she proposed that the amount of information should be adjustable by the participants themselves, allowing for a customized experience based on individual needs and preferences.

\section{\pathfinder}\label{sec:pathfinder}
\begin{figure*}
\centering
 \includegraphics[width=0.7\textwidth]{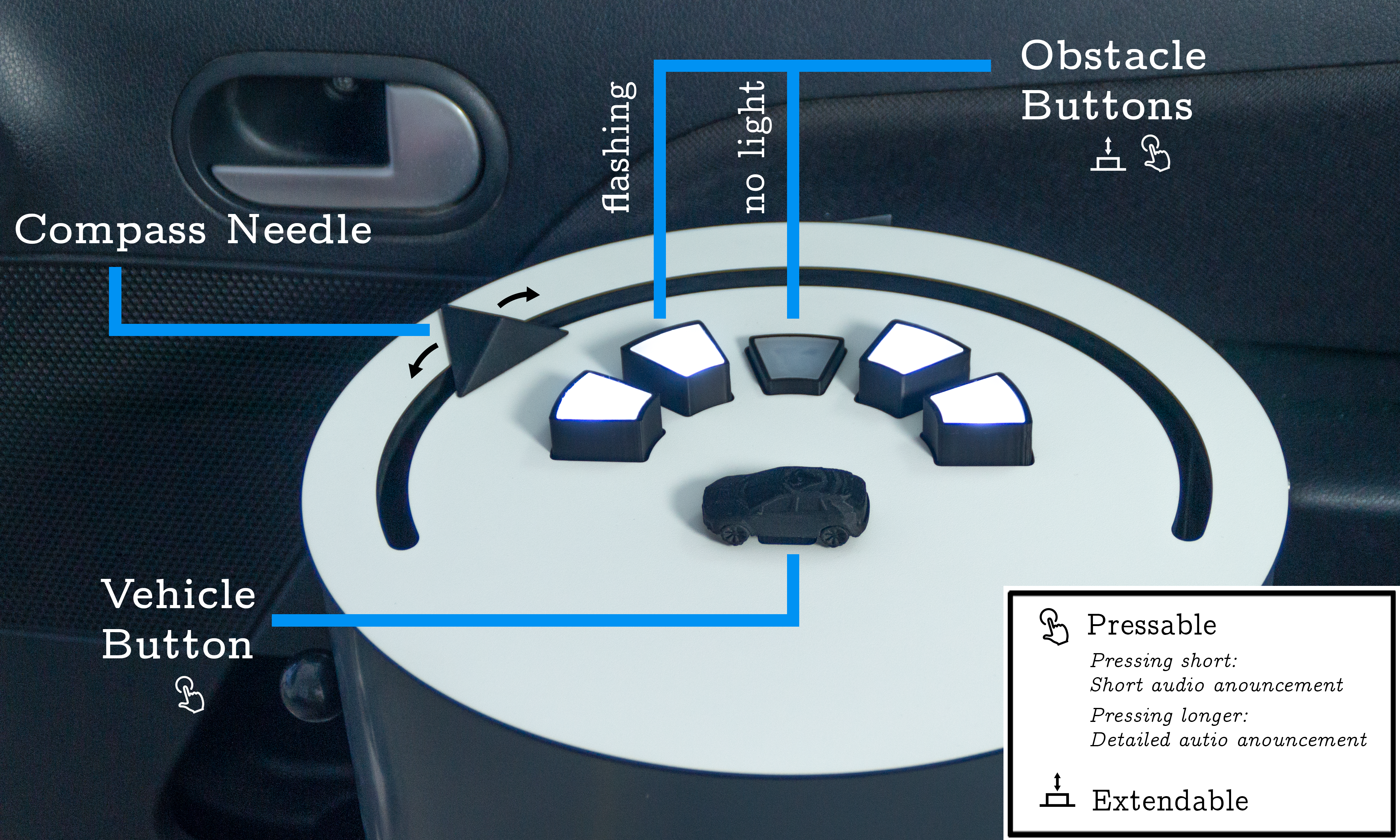}
  \caption{Interface design of \pathfinder, featuring a compass needle, five obstacle buttons, and a vehicle button. Each obstacle button can extend to indicate an obstacle in the corresponding section, and pressing it triggers additional details via audio announcements. The extended buttons also flash to enhance visibility. The compass needle moves along a rail to continuously point toward the final destination.}
  \Description{This photo shows the design of PathFinder with descriptions regarding the functionality of each element, such as the compass needle, the flashing five obstacle buttons and the vehicle button}
  \label{fig:interface_design}
\end{figure*}

Based on the insights of the interactive workshop, we developed \pathfinder, a multimodal interface that considers the individual needs of BVIPs and assists them in exiting HAVs (see \autoref{fig:interface_design}). This section will describe the design rationale and features of \pathfinder.

\subsection{General Design}
In general, our participants preferred combining tactile feedback with auditory cues. Consequently, tactile feedback should provide a broad overview of the surrounding environment, serving as a foundational layer of information, while verbal feedback adds detailed information.
Additionally, \pathfinder employs a clean and simple interface design to prevent users from becoming overwhelmed and used high contrasts in color (black and white) to enhance visibility for those with residual vision, as suggested in prior work exploring accessible technology for BVIPs~\cite{Meinhardt.2024, giudice2008blind, Holloway.2019}. As described below, \pathfinder consists of four main components positioned on an oval-shaped plate (30x24cm): the \textit{initial audio announcement}, the \textit{compass needle}, the \textit{five obstacle buttons}, and the \textit{vehicle button}. All electronic components of \pathfinder were controlled by an Arduino Mega microcontroller~\cite{arduino} and powered by an external power source. The following subsection gives a brief overview of the construction of each component. In addition, to reproduce \pathfinder, we have provided all the construction files, including blueprints, 3D files, and laser-cut files, in a git repository (see section: \nameref{sec:openscience}).

\subsection{Initial Audio Announcement}\label{sec:initialaudioanouncement}
To clearly distinguish between the HAV stopping at a traffic light~\cite{Meinhardt.2024} and the destination, we created an initial audio announcement that indicates that the HAV has parked at the destination. This announcement also provides directions and distances for the passengers to reach their final destination (e.g., a coffee shop) and informs them if pedestrians or cyclists are expected to pass by. A detailed description of the audio announcements can be found in \autoref{app:audio_announcements}. This verbal audio is played automatically once the HAV stops. According to the participant's feedback, this audio announcement is as concise as possible.

\subsection{Five Obstacle Buttons}
To represent the vehicle's surroundings, we divided the area into five sectors corresponding to the direction of exit. These sectors are represented by five buttons that can extend following the initial audio announcement. If a static obstacle is detected in a particular sector, the corresponding button extends; otherwise, it remains retracted. For blind participants, this difference in height can be sensed by touch. For those with residual vision, the extended buttons also blink to attract attention, as \citet{holloway_animations_2022} recommended to ``\textit{use blinking pins to direct attention to important areas [...]}''\cite[p. 12]{holloway_animations_2022}. This approach was also supported by \citet{Ivanchev_pre-journey_2014}, who discovered that blinking interactive elements were beneficial for navigation tasks among BVIPs.

Based on participants' feedback, we enhanced the tactile feedback system by adding audio announcements. We made all five buttons pressable, regardless of whether they were extended or retracted. Due to the resistance in our design, pressing a button does not cause it to move; instead, it maintains its current state. A short press provides concise information, while a long press delivers detailed information, including distances to the respective obstacles. This dual approach was implemented to meet the participants' ambiguity in the need for comprehensive but concise information. It also ensures that participants, including those with no remaining vision, have access to detailed information.

Our participants preferred a structured and organized approach that ensured they received all relevant information, giving them confidence that nothing was missed. This aligns with the findings of \citet{Brewer.2018.Understanding}, whose participants emphasized the importance of an interface that provides feedback in a clear and organized manner. 
Thus, by pressing each button, they can gain a complete understanding of the vehicle's surroundings. Additionally, embedding audio announcements for sector-specific obstacles into the button allows participants to actively seek out information rather than passively receive it, as already highlighted by \citet{arditi2013user}. Nevertheless, crucial information, such as passing dynamic obstacles (e.g., cyclists or pedestrians), is automatically announced as they approach the HAV, as mentioned in our workshop.

The mechanism enabling the buttons to be extendable was achieved using a camshaft system powered by a servo motor for each button, which raises the button. At the tip of each button, a white LED was embedded beneath a frosted acrylic glass plate, providing the flashing.

\subsection{The Compass Needle}
The compass needle provides directional guidance toward the final destination both during the ride and once the HAV has stopped. This concept was inspired by feedback from participants in a workshop conducted by \citet{Brewer.2018.Understanding} and was further validated by our participants, who emphasized the importance of knowing the direction of the final destination and the side of the vehicle to exit.

We constructed the compass needle using a stepper motor that moves a timing belt and a metal sled. The 3D-printed compass needle is mounted on this sled, allowing it to traverse an oval-shaped rail covering approximately 180°. If the final destination is to the vehicle's left during the ride, the needle will point as far left as possible within its limited range of motion.

\subsection{Vehicle Button}
\citet{Meinhardt.2024} noted that participants required a reference point, such as the ego vehicle, to contextualize all other information and gain spatial orientation. Hence, we constructed a button shaped like a vehicle as a reference point to the five obstacle buttons and the compass needle. Further, by pressing this button, the initial audio announcement (see \autoref{sec:initialaudioanouncement}) could be repeated to ensure participants could actively seek the information~\cite{arditi2013user}. Like the obstacle buttons, a detailed announcement would be played if pressed longer, including information about the terrain perturbations.

\section{User Study}
To investigate the capabilities of \pathfinder in assisting BVIPs to exit HAVs, we conducted a user study with N=16 BVIPs. We compared this multimodal interface with an auditory baseline inspired by the ATLAS system from \citet{Brinkley.2019}, chosen because the auditory output is currently the predominant modality used in interfaces for BVIPs~\cite{Chanana.2017}. The auditory cues provided comprehensive environmental information equivalent to that offered by \pathfinder, ensuring a fair comparison by delivering all detailed information upon the vehicle's arrival at its intended destination. The detailed description of the audio announcement of \pathfinder and the auditory baseline, including their durations, can be found in \autoref{app:audio_announcements}.

To enhance the generalizability of our study, we assessed both the auditory baseline and \pathfinder systems in two distinct scenarios: a complex urban environment and a simpler rural one. Based on findings by \citet{Meinhardt.2024}, which indicate that visual acuity affects how BVIPs engage with a system, we included visual acuity as a factor in our analysis. This led to a three-factor design in our study: \textit{System} (auditory/\pathfinder) and \textit{Scenario} (urban/rural) were the within-subject factors, while visual acuity was the between-subject factor.

The participants of the user study partly overlapped with the participants of the prior interactive workshop, as all workshop participants also took part in the user study (indicated in \autoref{app:demogr}). Their average age was \m{59.06}, \sd{15.00} (nine female, seven male and no non-binary). Their visual accuracy varied from total blindness (0\%) to 14\% with \m{4.88}, \sd{4.67}. Detailed information about the visual impairment of each participant can be found in \autoref{app:demogr}.

\subsection{Study Setup}
Aiming for high realism in our study, we utilized three 55" monitors positioned side by side to simulate the surroundings of the HAV, as illustrated in \autoref{fig:teaser}. For participants with residual vision, this setup provided a 180° view of the right side of the HAV's surroundings. The scenarios were created using Unity~\cite{unitygameengine} version 2020.3.15f2, incorporating various Unity assets (e.g.,~\cite{uts, naturePack, modernCity}). 

\begin{figure*}[ht!]
\centering
\small
    \begin{subfigure}[c]{0.49\linewidth}
        \includegraphics[width=\linewidth]{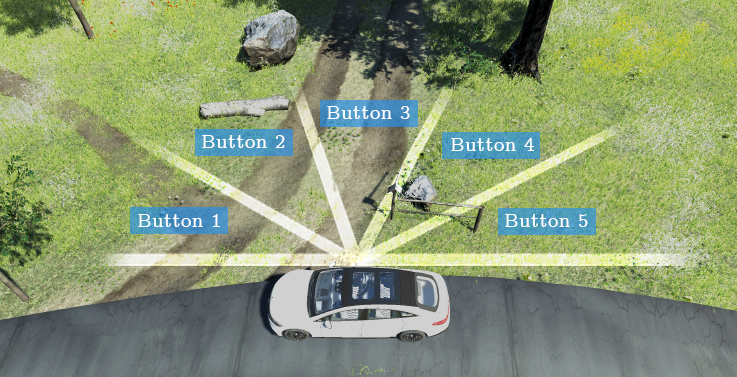}
        \caption{Sectors of the Obstacle Buttons in the Rural Scenario}
        \Description{Sectors of the five obstacle Buttons in the Urban Scenario from a top view}
    \end{subfigure}
    \hspace{0.6em}
        \vspace{0.6em}
    \begin{subfigure}[c]{0.49\linewidth}
        \includegraphics[width=\linewidth]{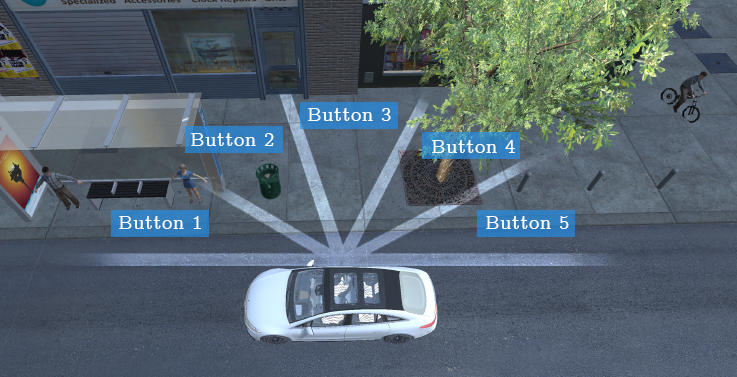}
        \caption{Sectors of the Obstacle Buttons in the Urban Scenario}
        \Description{Sectors of the Obstacle Buttons in the Urban Scenario from a Top View}
    \end{subfigure}

    \begin{subfigure}[c]{0.49\linewidth}
        \includegraphics[width=\linewidth]{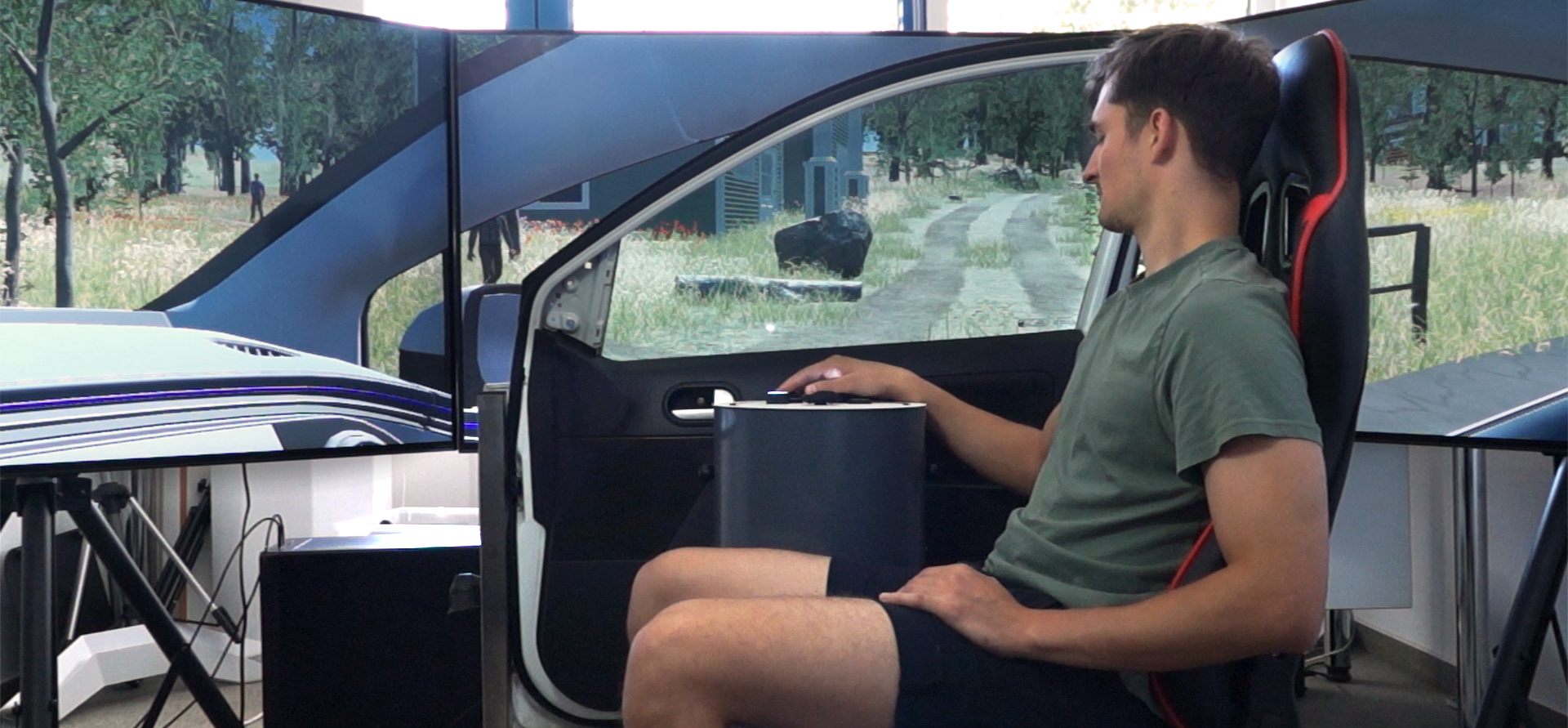}
        \caption{Participant's perspective of the Rural scenario}
        \Description{Study setup form rural scenario form a side view with a participant in the car seat}
    \end{subfigure}
    \hspace{0.6em}
    \begin{subfigure}[c]{0.49\linewidth}
        \includegraphics[width=\linewidth]{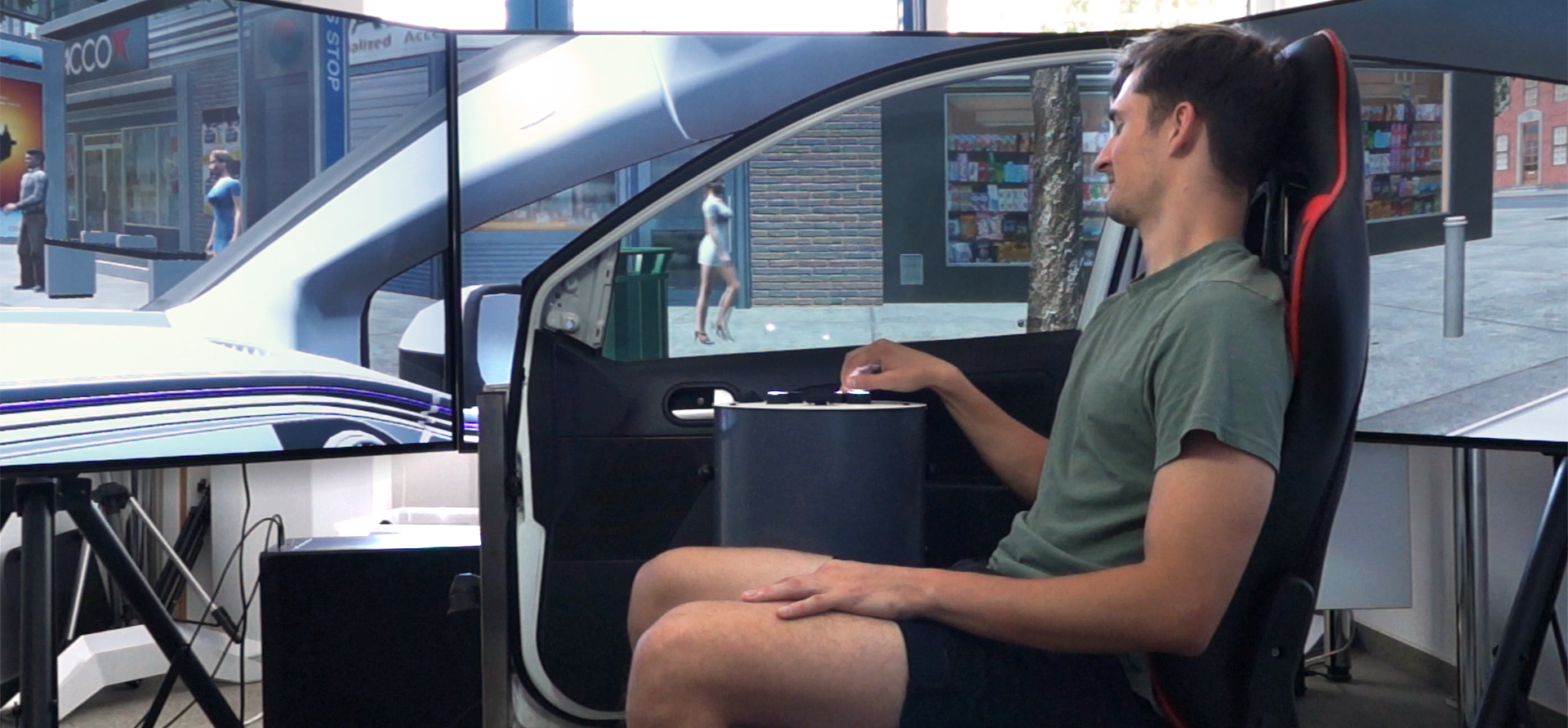}
        \caption{Participant's perspective of the Urban scenario}
        \Description{Study setup form urban scenario form a side view with a participant in the car seat}
    \end{subfigure}
   \caption{Sectors for the five Obstacle Buttons of \pathfinder for the Urban and Rural Scenario and the Participants' Perspective of the Study Setup. See \autoref{app:audio_announcements} for the concrete audio announcements}~\label{fig:szenarios}
   \Description{The figures display a top-down view of the HAV in both urban and rural scenarios, along with the corresponding study settings for each scenario. The urban scenario illustrates the HAV in a city environment, with buildings, streets, and pedestrians/cyclists. The rural scenario shows the HAV in a more open, less densely populated area, with fields, forests and sparse buildings}
\end{figure*}

Both scenarios featured simulated pedestrians and cyclists passing by the vehicle, with different complexities reflecting typical environmental variations. Hence, dynamic obstacles were more frequent in the urban scenario, averaging four pedestrians and two cyclists per minute. In contrast, the rural scenario averaged two pedestrians and one cyclist per minute. Additionally, obstacles in the urban scenario were distributed across four sections, while the rural scenario had them in two sections. The atmospheric audio also differed: the urban scenario featured bustling city sounds, including passing vehicles and people talking, while the rural scenario had a quieter ambiance with forest sounds, such as birds singing.
The vehicle's surroundings for the urban and rural scenarios can be seen in \autoref{fig:szenarios}. Further, the corresponding audio announcements for each obstacle button can be found in \autoref{app:audio_announcements}.

We reused the car door and seat used in the workshop (see \autoref{sec:carwindowprototype}) and positioned this setup in front of the center monitor. We ensured alignment with the virtual car window in the simulation from the passenger's point of view.
Further, we mounted \pathfinder between the car seat and the window, speculating a plausible position for this kind of future interface.
The study setup also included a camera facing the participant and a microphone to record qualitative feedback.

\subsection{Procedure}\label{sec:procedure_userstudy}
For each participant, we described the study setup in the room and sought the participants' consent to record the session. We ensured they comprehended all aspects and encouraged them to ask questions. We then read the consent form aloud, adhering to the research institute's ethical guidelines, highlighting their right to withdraw from the study at any time. The procedure also guaranteed privacy protection, anonymity, fair compensation, and risk aversion. Acknowledging the unique needs of our participants, we went beyond standard ethical practices by offering personalized support, such as assistance with transportation, to maintaining high ethical standards in accessibility research.

After obtaining their consent, the BVIPs were seated next to the car door and asked to imagine themselves as passengers in an HAV traveling to their desired destination without any need for intervention (SAE level 4 to 5)~\cite{SAElevel}. Before starting the four main conditions (i.e., urban and rural scenario, auditory baseline, and \pathfinder), participants were introduced to the study procedure through an introductory suburban scenario, where the vehicle drove for 10~seconds before reaching its destination. During this scene, we explained the functionalities of \pathfinder and the auditory cues while asking them to understand the surroundings of the surroundings. Participants were encouraged to repeat this introductory scenario as often as necessary to explore the interfaces until they felt comfortable with their features. While most participants completed the introductory scenario once, four participants requested to repeat it a second time. This introductory scenario was entirely different from the main scenarios to prevent any overlap of information and bias towards the main scenarios.

The four main conditions were then presented in a counterbalanced order. Each scenario included a 30-second ride before the HAV reached its destination, after which participants were told they had as much time as they needed to explore the vehicle's surroundings as best as they could using either \pathfinder or the auditory-only baseline. For the auditory condition, \pathfinder was covered with a wooden lid to prevent interaction with the system. Participants were also allowed to repeat the audio announcements as often as they wished. 
The simulation concluded once participants indicated they had obtained sufficient information to exit the vehicle. Notably, participants did not physically open the car door during the simulation.

For the urban scenario, participants were informed that the HAV would take them to a coffee shop in an unfamiliar area. In the rural scenario, they were told that the HAV would transport them to a friend's house, also in an unfamiliar area. After completing the four conditions, we collected demographic information, including age, gender, and visual acuity. We then engaged in a qualitative conversation asking the participants to compare both interfaces in relation to the insights from the workshop (see \autoref{sec:informationNeeds} and \autoref{sec:information_conveance}). Specifically, we asked about the clarity in conveying dynamic and static obstacles, the conveyance of terrain perturbations, the information provided about the final destination, the spatial orientation, and the overall completeness of information necessary for a comprehensive understanding of the HAV's surroundings.

The participants were compensated for the 1.5h session with 18 \anon{Euros}.

\subsection{Measurements}
After each condition, participants were asked to rate both \pathfinder and the auditory baseline as experienced within the respective scenario. We utilized the System Usability Scale (SUS)~\cite{SUS_questionaire} to assess usability. Additionally, we measured the participants' mental demand using the NASA-TLX scale~\cite{hart1988development}. To assess perceived situation awareness~\cite{endsley1998comparative, endsleyTheorySituationAwareness1995}, we employed the Situation Awareness Rating Technique (SART)~\cite{taylor2017situational}. We also evaluated the participants' perceived safety through a set of four 7-point semantic differential scales, ranging from -3 (anxious/agitated/unsafe/timid) to +3 (relaxed/calm/safe/confident)~\cite{faas2020longitudinal}. \\
Finally, we used the Immersion subscale of the Technology Usage Inventory (TUI)~\cite{kothgassner2013technology} to ensure that participants were sufficiently immersed during the study. This measurement helps us determine if the study’s findings are comparable to those in a potential real-world scenario.
All questionnaires were read aloud to ensure they were accessible to all participants.

\subsection{Results}
During our user study, we collected qualitative and quantitative results, which will be reported in the following two sections. After all conditions of the user study, participants rated their perceived immersion via the TUI~\cite{kothgassner2013technology} during the simulation as medium-high \m{17.06}, \sd{6.16} (minimum: 4, maximum: 28), indicating a reasonable approximation to potential real-world scenarios. On average, the time between the HAV stopping at the destination and participants indicating that they had sufficient information to exit the vehicle was 1~min 46~sec (\sd{59~sec}) for \pathfinder and 1~min 57~sec (\sd{55~sec}) for the auditory baseline. Refer to \autoref{app:descriptive_data_user_study} for detailed descriptive data.

\subsubsection{Quantitative Data}
To ensure our quantitative data met the assumptions necessary for statistical analysis, we first used the Shapiro-Wilk test~\cite{Shapiro-Wilk} to check for normality. For data that followed a normal distribution, we performed a repeated measures ANOVA. When the data did not meet the normality assumption, we applied the aligned rank transformation (ART) method, which is suited for non-parametric factorial analysis of repeated measures~\cite{wobbrock_aligned_2011}. The WHO categorizes visually impaired individuals into two groups: \textit{legally blind} and \textit{visually impaired}, with visual acuity of 5\% or less classified as \textit{legally blind}~\cite{levelofblindness}. Following this approach, we categorized participants into these two groups due to the limited data range available for participants' visual acuity. In our analysis, the system and scenario were treated as within-subject factors, while BVIPs' visual acuity was treated as the between-subject factor. This categorization resulted in ten participants being classified as \textit{legally blind}, while the other six participants were categorized as \textit{visually impaired}. 
We conducted our analyses using R software version 4.4.1.

\begin{figure}[ht!]
\centering
\small
    \begin{subfigure}[c]{0.48\linewidth}
        \includegraphics[width=\linewidth]{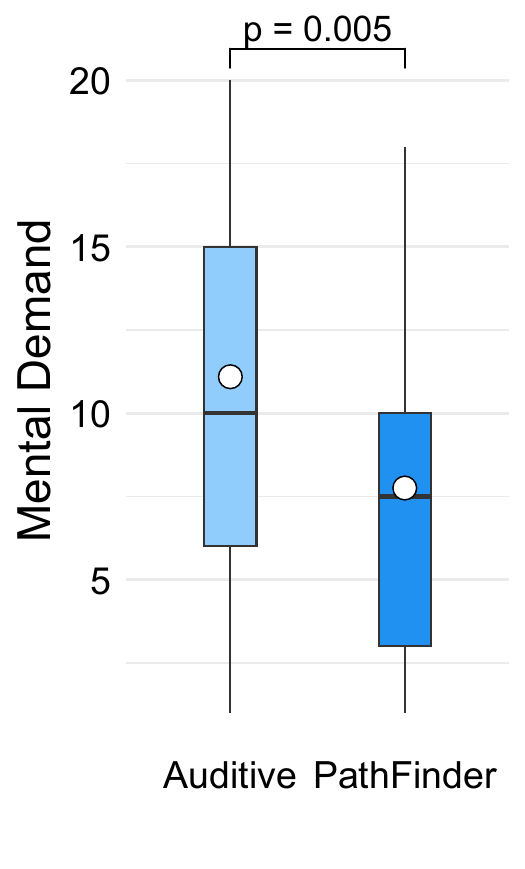}
        \caption{Significant main effect on mental demand~\cite{hart1988development} for \textit{System}}
        \label{fig:mental_demand}
        \Description{The figure presents a bar plot illustrating the main effect on mental demand for the system, showing that the auditory-only baseline has a significantly higher mental demand compared to PathFinder}
    \end{subfigure}
    \hspace{2mm}
    \begin{subfigure}[c]{0.48\linewidth}
        \includegraphics[width=\linewidth]{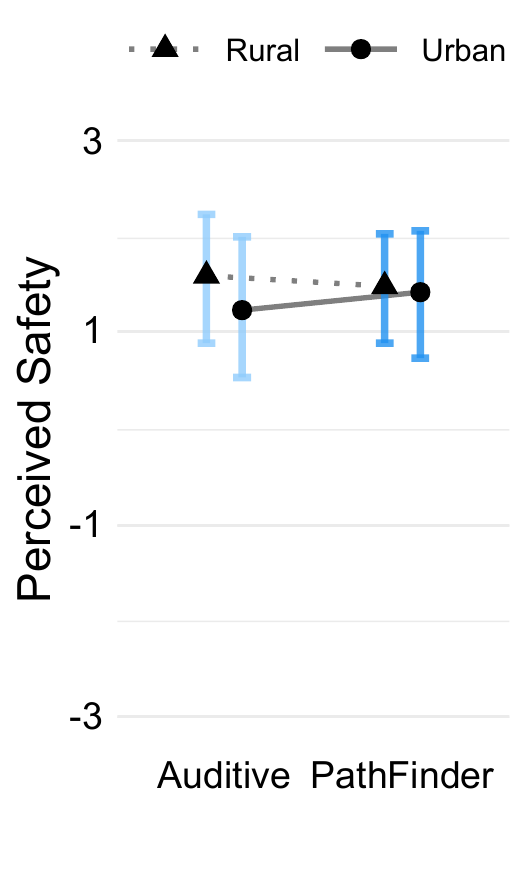}
        \caption{Significant interaction effect on perceived safety~\cite{faas2020longitudinal}}
        \label{fig:perceived_safety}
        \Description{This figure presents a line plot showing the interaction effect of Scene and System}
    \end{subfigure}
   \caption{Significant effects of the quantitative data for mental demand and perceived safety}~\label{fig:quant_results}
   \Description{This figure presents significant effects of the quantitative data collected, showing the impact on mental demand and perceived safety across the two systems and scenarios.}
\end{figure}

\paragraph{Mental Demand}
The ART found a large significant main effect on mental demand~\cite{hart1988development} for \textit{System} (F(1) = 11.03, p = 0.005; $\eta^2$ = 0.44, 95\% CI [0.11, 1.00]). Hence, \pathfinder (\m{7.75}, \sd{5.16}) yielded a significantly lower mental demand than the auditory Baseline (\m{11.09}, \sd{5.63}). Refer to \autoref{fig:mental_demand} for the plotting. 
Further, the ART found a trend towards significance, suggesting an interaction between \textit{Scenario} and \textit{Visual Acuity} (p=0.063). While not significant, the interaction's $\eta^2$ = 0.23 effect size is defined as large~\cite{cohen_statistical_2013}.

\paragraph{Usability} The ART did not identify any significant main or interaction effects on Usability~\cite{SUS_questionaire}. The usability ratings for \pathfinder (\m{64.84}, \sd{16.30}) were similar to those for the Auditory Baseline (\m{62.42}, \sd{14.07}). 

\paragraph{Situation Awareness} An ANOVA did not find significant differences in situation awareness~\cite{taylor2017situational}. Yet, the rated situation awareness for both \pathfinder (\m{19.09}, \sd{5.51}) and the auditory baseline (\m{17.91}, \sd{6.25}) was rated medium to high on a scale from -20 to 40.

\paragraph{Perceived Safety} The ART found a large significant interaction effect between \textit{System} and \textit{Scenario} (F(1) = 6.47, p = 0.023; $\eta^2$ = 0.32, 95\% CI [0.03, 1.00]). Hence, Perceived Safety ratings for \pathfinder were consistent across the Urban (\m{1.44}, \sd{1.86}) and Rural scenarios (\m{1.50}, \sd{2.00}). In contrast, the ratings for the auditory baseline showed a divergence between the urban (\m{1.25}, \sd{1.25}) and rural (\m{1.61}, \sd{2.13}) scenario. No main effects were found for either \textit{System}, \textit{Scenario}, or \textit{Visual Acuity}. Refer to \autoref{fig:perceived_safety} for the plotting.

\subsubsection{Qualitative Feedback}
After completing all four conditions, we conducted brief interviews with participants, focusing on their experiences with both interfaces in relation to the findings from the previous workshop (see \autoref{sec:informationNeeds} and \autoref{sec:information_conveance}). Unlike the workshop, we did not perform a thematic analysis of this feedback. Instead, we present anecdotal feedback and participants' opinions, organized by the specific questions asked during the interviews. Therefore, the clusters presented below are based directly on the specific questions (see \autoref{sec:procedure_userstudy}) asked during the interviews rather than being the result of a formal thematic analysis.

\textbf{Spatial Orientation.}
The feedback on spatial orientation varied across participants, but the majority of participants were satisfied with the conveyed information of \pathfinder with regards to the spatial orientation. \PoneStudy found the \pathfinder particularly compelling, stating, ``\textit{I was impressed because I could get an overview by pressing different symbols}''. This opinion was echoed by \PfiveStudy, who also preferred the multimodal interface, noting that it allowed her to ``\textit{create a mental map}'', whereas listening to the auditory feedback led to higher mental demand. However, \PthreeStudy expressed some confusion, stating that ``\textit{using ‘left’ and ‘right’ instead of ‘in the direction of travel’ was disorienting}''.
\PfifteenStudy indicated that he was able to achieve sufficient spatial orientation through the auditory baseline, whereas the tactile components of \pathfinder proved particularly distracting, necessitating concentration on the individual elements, which in turn impeded his ability to listen to the audio announcements carefully. 

\textbf{Navigation to the Final Destination}
Four participants stated that the information how to navigate to the final destination was very similar between both interfaces. Particularly, \PeightStudy mentioned that ``\textit{the information was the same but the method was different}''. \PtwoStudy appreciated the tactile compass needle of \pathfinder to navigate to the final destination, stating, ``\textit{The triangle [compass needle] in the multimodal interface was better for me. I knew where I was and where I needed to go}''. Conversely, \PfifteenStudy stated that the compass needle was not needed as the initial audio announcement was already sufficient to navigate to the final destination.
However, \PfiveStudy found the navigation through touch more intuitive than listening to the auditory baseline. But she appreciated the combination of both interfaces, saying, ``\textit{Both should be combined, but I prefer touch for navigation}''. 

\textbf{Terrain Perturbations.}
Most participants felt that the information about the terrain perturbations was clear and sufficient across both interfaces. For instance, \PfourStudy mentioned that ``\textit{it was clear whether the ground was paved or not}''.  However, \PfifteenStudy criticized that with \pathfinder, he had to search for terrain information, whereas the auditory baseline provided it automatically. Further, \PelevenStudy noted that some information was excessive, such as the grass on the sidewalk.

\textbf{Dynamic and Static Obstacles.}
Participants generally agreed that the identification of dynamic and static objects was essential but differed in their preferred modality of receiving this information. \PoneStudy preferred \pathfinder for recalling details of the static obstacles. \PtwoStudy and \PthreeStudy emphasized the importance of combining both interfaces, with \PtwoStudy suggesting, ``\textit{A pin that rises when a pedestrian is present would be helpful}.'' This suggestion is particularly notable, as the workshop's findings indicated that dynamic obstacle information was preferred to be conveyed verbally for quicker understanding.

\PeightStudy and \PtenStudy highlighted the challenge of predicting the presence of dynamic objects like pedestrians or cyclists. They noted the absence of information indicating when these obstacles had passed, which would signal that it is safe to exit the HAV. However, \PsixteenStudy appreciated that both interfaces announced the presence of these obstacles, stating that knowing they are nearby is more important to her than precisely when they pass by the vehicle. \PsevenStudy also voiced concern about the lack of continuous updates on dynamic objects, expressing a desire for a system that ``\textit{always informs me when the situation changes}.''

\textbf{Completeness of Information.}
Overall, participants felt that both systems provided comprehensive information, though seven participants, including \PsixStudy, \PeightStudy, and \PnineStudy, mentioned that the auditory information was overwhelming, with \PnineStudy specifically stating, ``\textit{It says too much, and I have to concentrate hard}''. \PsevenStudy criticized that both interfaces only covered the area immediately around the HAV, leaving users without further guidance once they moved beyond a few meters. However, this concern contrasts with insights from the prior workshop, where participants preferred using traditional mobility aids, such as canes or guide dogs, after their immediate situation awareness needs were satisfied (see \autoref{sec:informationNeeds}).

\section{Discussion}
This research was driven by the need for BVIPs to gain assistance when exiting HAVs in unfamiliar environments~\cite{Brewer.2020}. In an interactive workshop (N=5), we found that BVIPs currently rely on acquaintances to gain situation awareness of the vehicle's surroundings. However, with the introduction of HAVs, BVIPs may gain more independence~\cite{Kacperski.2024}, but they likely face situations alone without human assistance. To investigate the information needs of BVIPs when exiting HAVs, we presented three low-fidelity prototypes to the participants. Feedback from the workshop indicated a preference for a multimodal approach to convey information about the environment in an organized and structured manner. Based on this feedback, we developed \pathfinder, which integrates visual, tactile, and auditory cues to assist BVIPs. Using the Participatory Design approach~\cite{ParticipatoryDesign.1993}, we involved BVIPs from the outset, ensuring that \pathfinder's final design met the diverse needs of users with varying degrees of visual impairment. This approach aligns with the recommendations of \citet{bradley_experimental_2005} and \citet{albouys-perrois_towards_2018}, who emphasized the importance of designing audio and tactile cues based on specific user needs and preferences for navigational tools for BVIPs.
We subsequently conducted a three-factorial within-between-subject user study (N=16), simulating an HAV ride. Our study assessed \pathfinder against an auditory-only baseline in both complex urban and simpler rural scenarios. \pathfinder yielded a significantly lower mental demand than the auditory baseline and maintained high perceived safety in both scenarios, while the auditory baseline led to lower perceived safety in urban scenarios compared to rural ones. 

% \subsection{Integrating User Feedback into the Design Process}
% Using the Participatory Design approach~\cite{ParticipatoryDesign.1993} allowed us to involve BVIPs from the beginning of the design process, ensuring that \pathfinder's final design met the diverse needs of users with varying degrees of visual impairment. The workshop findings were fundamental in designing and constructing \pathfinder, particularly in identifying the need for combining auditory, tactile, and visual feedback to enhance situational awareness. The final user study validated these design choices, demonstrating that the multimodal prototype significantly reduced mental demand compared to an auditory-only baseline.

% This outcome underscores the importance of integrating BVIPs' feedback early and continuously throughout the design process, as already suggested by \citet{bradley_experimental_2005}. Hence, we advocate for the broader adoption of this approach in developing accessible systems, as it ensures that the final designs are closely aligned with the real needs and preferences of the intended users.

\subsection{Multimodal Approaches to Convey Environmental Information}
Our findings indicate that the multimodal \pathfinder interface is effective in conveying information about the HAV's surroundings, enabling participants to create "mental maps" and gain situation awareness. This finding is important because developing accurate cognitive maps of the transportation environment is essential for BVIP independence and mobility~\cite{palani_design_2020, Fink.2023.2, fink_expanded_2023} and aligns with broader evidence supporting the effectiveness of multimodal interfaces; for instance, \citet{papadopoulos_cognitive_2017} highlighted that audio-tactile maps enhance BVIP's spatial orientation, especially in unfamiliar environments. Further, the reduced mental demand found for \pathfinder supports the \textit{multiple-resource theory}~\cite{wickens_compatibility_1983}, which posits that cross-modal distribution of information reduces competition for cognitive resources, thereby reducing mental demand. This aligns with participants' statements, appreciating the combination of the modalities, especially the auditory and tactile ones. The broad implications of our multimodal approach to improving mental mapping and reducing mental demands can be realized in efforts to increase independent mobility for BVIP passengers. Just as our workshop participants reported that they often rely on drivers or acquaintances to help them understand the environment and exit safely, it stands to reason that interfaces like \pathfinder can help improve safety and independent travel for BVIPs in future HAV.       

Interestingly, while our qualitative results clearly prefer \pathfinder over the auditory baseline concerning spatial orientation, the quantitative analysis revealed no significant differences in situation awareness between the two systems. This discrepancy may arise from participants' challenges in accurately self-reporting their situation awareness, as suggested by \citet{Endsley.1998.comapre}, or because both systems convey the same information, resulting in similar situation awareness ratings. The latter aligns with the fact that both systems received medium to high ratings for situation awareness, suggesting that they are generally effective in this regard. These findings, however, diverge from those of \citet{Meinhardt.2024} and \citet{Md.Yusof.2020}, who reported low ratings for situation awareness with their tactile interfaces. Yet, it is important to note that their studies focused on conveying traffic information during the HAV ride, whereas our study centered on the HAV's surroundings when exiting the vehicle. This difference in the journey's parts is interesting, as understanding the HAV's surroundings when exiting into an unfamiliar area is likely more critical for BVIPs than being immediately aware of the traffic situation during the ride. While situation awareness during the journey is also important~\cite{Meinhardt.2024, Fink.2023.2}, it becomes essential when navigating a new environment after exiting the vehicle. This difference in context might explain the variation in situation awareness ratings across different studies.

\subsection{Inconsistency in Diverse Scenario's Complexities}
Our study revealed a significant interaction effect between the scenarios and the two systems on perceived safety (see \autoref{fig:perceived_safety}). While we expected that the more complex urban scenario would lead to differing ratings, \pathfinder consistently maintained high perceived safety across both urban and rural settings. In contrast, the auditory baseline showed divergence, with lower perceived safety in the urban scenario compared to the rural one. This indicates that while the auditory baseline may meet BVIPs' safety needs in simpler environments, it becomes less reliable in more complex settings. 
These findings underscore the limitations of single-modality approaches~\cite{Yatani.2012, Fink.2023.2} and suggest that multimodal systems like \pathfinder offer greater robustness across varying levels of environmental complexity. Therefore, these results support the recommendations of \citet{kuriakose_tools_2022}, who highlight that multimodal cues enable BVIPs to adjust their information intake according to situation demands.
This robustness is essential for ensuring safety in demanding scenarios where situation awareness is critical. However, it is important to recognize that our study was limited to only two scenarios. While these scenarios were designed to reflect typical environments BVIPs might encounter, they do not capture the full range of possible conditions that could affect information conveyance. For instance, extreme weather conditions such as heavy rain, snow, or fog could introduce new challenges that neither the multimodal nor the auditory-only system might handle effectively. Further, we used simple scenarios, implying that after the initial obstacles, the path to the final destination is straight. While this might not reflect real-world scenarios, we based this decision on the participants' statement that after exiting the vehicle, they would rely on traditional mobility aids such as canes or guide dogs (see \autoref{sec:informationNeeds}). This decision reflects the interface's primary purpose: providing essential initial information conveyance before users switch to their customary navigation methods. However, in scenarios with no obstacles, the auditory-only system might suffice and even be preferred due to its simplicity.

\subsection{One System to Rule the Entire Journey}\label{sec:one_system_to_rule}
This research contributes to the growing body of work on exploring accessible interfaces for each part of a journey using HAVs, such as finding the vehicle~\cite{fink_multisensory_2024} or conveying information during the ride~\cite{Fink.2023.2, Meinhardt.2024, Md.Yusof.2020}. We extend this work by focusing specifically on the crucial exiting phase. 
Previous studies~\cite{Fink.2023.3, giudice2008blind} have highlighted BVIP frustration with using multiple apps and systems for different navigation tasks. To address this problem, \citet{giudice2008blind} suggested that integrating systems could enhance effectiveness across different scenarios. 

Therefore, it seems desirable to combine the \pathfinder system with other tactile or multimodal systems (e.g.,~\cite{Fink.2023.2, Meinhardt.2024}) to ensure comprehensive accessibility throughout the entire journey. However, integrating multiple functionalities into a single system requires careful consideration of the form factor to maintain ease of use. For example, simplifying \pathfinder by removing the compass needle—-considered unnecessary by participants-—can help reduce its size. Additionally, leveraging existing devices like smartphones~\cite{Fink.2023.3, Fink_AUto_UI} and tablets~\cite{Brinkley.2019} can extend the system's capabilities without increasing its size. For instance, smartphones could provide additional vibrotactile feedback on the HAV's location on a map~\cite{Fink_AUto_UI}. Our workshop findings indicate that BVIPs prefer systems that deliver essential information upfront, allowing them to keep their hands free for tasks such as using canes or guide dogs. Therefore, expanding \pathfinder with the tactile elements on OnBoard~\cite{Meinhardt.2024}, like the rotating vehicle representation and the reason-for-stopping button, could enhance the user's understanding of the ongoing traffic during the ride.

\subsection{Practical Implications and Future Work}
While the quantitative data from our user study shows significant effects on mental demand and perceived safety, there are no significant differences between \pathfinder and the auditory-only baseline regarding usability and situation awareness. This suggests that an auditory-only solution may be sufficient for enhancing the exiting phase for BVIPs, potentially reducing the cost and complexity of adding tactile modality to the system.
However, qualitative feedback from participants highlights that for optimal effectiveness, information should be conveyed through all available modalities. For example, \pathfinder communicated dynamic obstacles only via audio. Yet, to enhance redundancy across modalities, these obstacles could also be conveyed using a tactile approach, such as bars that rise to indicate the presence of cyclists or pedestrians. This would further improve the system's robustness, providing that critical information is reliably understood by all potential passengers, regardless of their sensory preferences or extent, etiology, or onset of visual impairment. Furthermore, the significantly reduced mental demand observed with \pathfinder, along with its consistently high perceived safety in complex and simple scenarios, highlights its potential as a valuable add-on feature for vehicle manufacturers committed to accessibility. Additionally, future research should look into a more seamless integration of the tactile elements of \pathfinder into vehicles, such as using textile buttons and sliders integrated directly into the vehicle's fabrics~\cite{Nowak.2022, Schaefer.2023} or the armrest close to the door handle. The other modalities of \pathfinder could also provide more detailed information, such as whether a dynamic obstacle is moving fast or slow, via audio or blinking the obstacle buttons in different colors to distinguish between different types of obstacles for those with residual vision.

Finally, conducting real-world testing in actual vehicle environments would be essential to validate the system's effectiveness outside of controlled settings, ensuring that \pathfinder meets the practical needs of BVIPs in everyday use.

\subsection{Limitations}
Our interactive workshop included only five BVIPs. While small sample sizes can still provide valuable insights~\cite{toner2009small}, there is a potential for response bias~\cite{ming_accept_2021}. Thus, it is important to recognize that the views expressed by these participants may not fully represent the broader target group. Additionally, the design of \pathfinder was partly influenced by the subjective opinions of these five participants. While related work informed the development of both the auditory interface and \pathfinder, these interfaces should be considered with caution. Another limitation is the lack of external validity in our study, as participants did not physically exit an actual vehicle, which may affect the applicability of our findings to real-world scenarios. The actual process of exiting can introduce additional challenges, such as maintaining orientation, managing personal belongings or guide dogs, and navigating immediate hazards outside the vehicle. Additionally, testing the interfaces in a controlled environment rather than a real vehicle may have reduced the perceived risk and mental demand associated with exiting in real traffic conditions. These factors might have influenced participants' feedback and limit the generalizability of our results. Despite this, our study setup achieved a high level of perceived immersion, suggesting that the simulated experimental conditions were well-designed and effective. 

Additionally, due to the specialized nature of our target group, the user study was conducted with a relatively small sample size of N=16 for quantitative analysis. This sample size may limit the findings' applicability to a wider population. Moreover, fewer participants increase the risk of Type II errors, where true effects may not reach statistical significance. Therefore, it is important not only to consider statistical significance but also to examine the effect sizes. For example, although not statistically significant, the interaction between the \textit{Scenario} and \textit{Visual Acuity} on mental demand showed a large effect size. This suggests that there could be meaningful differences that warrant further investigation. Further, it is worth noting that while we attempted to provide a more nuanced perspective based on visual impairment (by highlighting qualitative responses with acuity information from \autoref{app:descriptive_data_user_study}) than the typical approach of collapsing BVIPs into a single group~\cite{Fink.2023.Chauffeur}, a larger sample size would have also enabled comparisons for the quantitative data. Additionally, whether participants were congenitally blind or acquired their impairment later in life could influence their specific information needs and should be explored in future studies.

It is also crucial to account for potential novelty effects~\cite{NoveltyEffect.1995} in our user study as the participants experience both interfaces for the first time. Hence, we anticipate that, as users become familiar with the interfaces over time~\cite{Mendoza.2005}, these novelty effects may diminish.  Specifically, the auditory baseline featured longer audio announcements compared to \pathfinder, which could have biased participants towards preferring \pathfinder. However, the similar time required for participants to gather sufficient information to exit the vehicle (see \autoref{app:descriptive_data_user_study}) suggests that the length of the auditory announcements did not impact the overall task performance.

Additionally, we were unable to counterbalance the between-factor of visual acuity, meaning that participants with similar visual acuity levels might have experienced the same order of conditions. This lack of counterbalancing could introduce slight learning effects, where participants become more accustomed to the tasks, potentially influencing the study’s results.

\section{Conclusion}
This paper introduces \pathfinder, a multimodal interface designed to assist BVIPs in safely exiting HAVs by providing information about the vehicle's surroundings. \pathfinder integrates visual, tactile, and auditory cues, making it accessible to users regardless of their visual impairment.

We conducted an interactive workshop with N=5 visually impaired participants to identify their information needs for safely exiting a vehicle. The workshop revealed that BVIPs currently rely heavily on acquaintances for assistance. However, as HAVs offer greater mobility independence, BVIPs may increasingly face these situations without human assistance. During the workshop, we presented three low-fidelity prototypes (a smartphone, a window touch prototype, and tactile bars), each employing different modalities and interaction strategies to assist with vehicle exit. Participants expressed a strong preference for a multimodal interface, favoring tactile cues as a foundation, supplemented by auditory cues for critical information, such as the presence of dynamic obstacles like cyclists. Based on these insights, we developed \pathfinder, a multimodal interface tailored to the unique needs of BVIPs. The system includes a compass needle that points to the final destination, five extendable, flashing obstacle buttons that represent different sections of the vehicle's surroundings and provide audio announcements for additional information, and a vehicle button that serves as a reference point.

In a subsequent three-factorial, within-between-subject user study (N=16), we evaluated \pathfinder against an auditory-only baseline in both complex urban and simpler rural scenarios. The results showed that \pathfinder significantly reduced mental demand compared to the baseline and consistently maintained high perceived safety in both scenarios. In contrast, the auditory baseline resulted in lower perceived safety in the urban scenario compared to the rural one. Further, the qualitative feedback indicated a clear preference for multimodal information conveyance to enhance spatial orientation and situation awareness. However, to increase robustness and ensure that critical information is reliably understood by all passengers, regardless of their sensory preferences or visual impairments, it is recommended that all information be conveyed across all modalities.

\section*{Open Science} \label{sec:openscience}
The source code and construction files, including blueprints, 3D-printing files, and laser-cutting files for both the three initial low-fidelity prototypes and \pathfinder have been made publicly available. These resources can be accessed at the following link: \\
\url{https://github.com/luca-maxim/light_my_way}.

\begin{acks}
This research was funded by the Deutsche Forschungsgemeinschaft (DFG, German Research Foundation) through the project “\textit{Non-Visual Interfaces to Enable the Accessibility of Highly Automated Vehicles for People with Vision Impairments}” (Project number: 536409562). 
The first author would like to thank Max Rädler for the support during the interactive workshop and Gertrud Vaas for being an esteemed contact person throughout this research journey. 

We also wish to acknowledge the \href{https://www.uni-ulm.de/en/einrichtungen/scientific-workshop/}{Scientific Workshop of Ulm University} for their assistance in constructing \pathfinder and the low-fidelity prototypes. Special thanks go to Alex Vogel, Wolfgang Rapp, and Manfred Kley for their practical support and dedication.

We also would like to thank Johannes Schöning for his valuable mental support during the \textit{CHITogether 2024~\cite{scott2024doing}} in St. Gallen.
\end{acks}

\bibliographystyle{ACM-Reference-Format}
\bibliography{sample-base}

\appendix

\section{Audio Announcements of \pathfinder and the Auditory Baseline}\label{app:audio_announcements}

Due to the participant's mother tongue, the audio announcements were in \anon{German}. For this appendix, we translated the audio announcements via \url{Deepl.com}.

\subsection{Urban Scene}

\subsubsection{\pathfinder.} 
\textbf{Upon Reaching the Destination,} the following audio announcement was played automatically: \\
"We have reached the end of the journey. The destination, Café Good Times, can be reached via the pavement next to the road. The entrance to the café is 70 meters away on the right-hand side." 
\textbf{Duration:} 0:14~min

\textbf{Vehicle Button Short}.
"The destination, Café Good Times, can be reached via the pavement next to the road. The entrance to the café is 70 meters away on the right."
% \smallskip

\textbf{Vehicle Button Detailed}.
"The destination, Café Good Times, can be reached via the pavement next to the road. The entrance to the café is 70 meters away on the right. The curb is one meter away from the vehicle. The pavement is 3 meters wide, paved, and level. There are many pedestrians and cyclists in front of the vehicle."
% \smallskip

\textbf{Obstacle Buttons}

\begin{itemize}[leftmargin=*, itemsep=0pt]
    \item \textbf{Button 1: Raised}
    \begin{itemize}
        \item \textit{Short}. "Bus stop."
        \item \textit{Detailed}. "Bus stop on the pavement five meters away."
    \end{itemize}
    \item \textbf{Button 2: Raised}
    \begin{itemize}
        \item \textit{Short}. "Rubbish bin."
        \item \textit{Detailed}. "Rubbish bin on the pavement three meters away."
    \end{itemize}
    \item \textbf{Button 3}
    \begin{itemize}
        \item \textit{Short}. "The pavement is clear."
        \item \textit{Detailed}. "The pavement is clear. The curb is one meter away."
    \end{itemize}
    \item \textbf{Button 4: Raised}
    \begin{itemize}
        \item \textit{Short}. "Tree."
        \item \textit{Detailed}. "Tree on the pavement three meters away. No danger from low-hanging branches."
    \end{itemize}
    \item \textbf{Button 5: Raised}
    \begin{itemize}
        \item \textit{Short}. "Three bollards."
        \item \textit{Detailed}. "Three bollards in a row on the pavement three meters away."
    \end{itemize}
\end{itemize}

\subsubsection{Auditory Baseline.} \textbf{Upon Reaching the Destination,} the following audio announcement was played automatically: \\
"We have reached the end of the journey. The destination, Café Good Times, can be reached via the pavement next to the road. The entrance to the café is 70 meters away on the right-hand side. The curb is one meter away from the vehicle. The pavement is three meters wide, paved, and level. There are many pedestrians and cyclists in front of the vehicle. There is a bus stop five meters to the left of the vehicle. A rubbish bin is three meters away from the bus stop. The pavement in front of the car door does not have obstacles. A tree is on the pavement three meters away to the right behind the vehicle. No danger from low-hanging branches. To the right of the tree, there are 3 bollards in a row on the pavement at a distance of 3 meters." 
\textbf{Duration:} 0:57~min

\subsection{Rural Scene}

\subsubsection{\pathfinder}
\textbf{Upon Reaching the Destination,} the following audio announcement was played automatically: \\
"We have reached the end of the journey. The destination, Carmen's House, can be reached via the dirt track right next to the road. The entrance is 15 meters straight ahead on the left-hand side." \\
\textbf{Duration:} 0:13~min
% \smallskip

\textbf{Vehicle Button Short}.
"The destination, Carmen's House, can be reached via the field path directly next to the road. The entrance is 15 meters straight ahead on the left-hand side."
% \smallskip

\textbf{Vehicle Button Detailed}.
"The destination, Carmen's House, can be reached via the field path directly next to the road. The entrance is 15 meters straight ahead on the left-hand side. The edge of the road is one meter from the vehicle. The field path is 2 meters wide, and the surface is unpaved and uneven. There are pedestrians and cyclists in front of the vehicle."
% \smallskip

\textbf{Obstacle Buttons}

\begin{itemize}[leftmargin=*, itemsep=0pt]
    \item \textbf{Button 1: Raised}
    \begin{itemize}
        \item \textit{Short}. "Trees"
        \item \textit{Detailed}. "Several trees next to the road on the grass five meters away."
    \end{itemize}
    \item \textbf{Button 2}
    \begin{itemize}
        \item \textit{Short}. "The area is clear."
        \item \textit{Detailed}. "The area is clear. A footpath crosses the field path three meters away."
    \end{itemize}
    \item \textbf{Button 3}
    \begin{itemize}
        \item \textit{Short}. "The field path is clear."
        \item \textit{Detailed}. "The field path is clear and consists of two channels. A branch and a stone lie on the grass next to the field path at a distance of seven meters."
    \end{itemize}
    \item \textbf{Button 4: Raised}
    \begin{itemize}
        \item \textit{Short}. "A fence and a stone behind it."
        \item \textit{Detailed}. "There is a fence on the grass next to the road two meters away. Behind it is a large stone three meters away."
    \end{itemize}
    \item \textbf{Button 5}
    \begin{itemize}
        \item \textit{Short}. "The area is clear."
        \item \textit{Detailed}. "The area is clear. The ground is a meadow."
    \end{itemize}
\end{itemize}

\subsubsection{Auditory Baseline.} \textbf{Upon Reaching the Destination,} the following audio announcement was played automatically: \\
"We have reached the end of the journey. The destination, Carmen's House, can be reached via the dirt track right next to the road. The entrance is 15 meters straight ahead on the left-hand side.
The edge of the road is one meter away from the vehicle. The dirt track is 2 meters wide and the surface is unpaved and uneven. There are pedestrians and cyclists in front of the vehicle.
There are several trees five meters to the left in front of the vehicle. Behind the trees is a path that crosses the dirt track 3 meters from the vehicle.
The country lane begins straight ahead. The track is clear and consists of two channels. A branch and a stone lie on the grass next to the field path at a distance of seven meters. 
To the right behind the vehicle is a fence next to the road on the grass two meters away. Behind it is a large stone three meters away." \\
\textbf{Duration:} 1:01~min

\onecolumn

\section{Participants' Demographic Data}\label{app:demogr}
The alpha level of blue highlighting of the participant IDs indicates their visual acuity in the tables below.
\begin{table*}[ht]
\caption{Table of participants' demographic data for the interactive workshop}
\Description{Table of participants' demographic data for the interactive workshop, including their age, gender, visual acuity, and a detailed description of their visual impairment as stated during the workshop}
\begin{tabularx}{\textwidth}{llllp{8cm}}
\label{tab:demografics_workshop}
\textbf{ID}     &\textbf{Age}   &\textbf{Gender} & \textbf{Visual Acuity} & \textbf{Impairment}  \\
\hline\hline
\Pone              & 61            & M      & 5\%           & total blindness on the left eye, right eye blurry vision  \\
\Ptwo              & 44            & F      & 1.5\%          & only contours visible  \\
\Pthree             & 52            & F      & 0\%           & total blindness \\
\Pfour             & 67            & F      & 0\%           & total blindness \\             
\Pfive   & 65  & M      & 3.5\%           & blurry vision  \\
\hline\hline
\end{tabularx} 
\end{table*}

\begin{table*}[ht]
\caption{Table of participants' demographic data from the user study and their overlapping participation with the workshop}
\Description{Table of participants' demographic data from the user study, including their age, gender, visual acuity, and a detailed description of their visual impairment as stated during the user study. Additionally the overlapping participants wit the prior workshop are indicated with an X}
\begin{tabularx}{\textwidth}{llllp{6.8cm}c}
\label{tab:demografics_userstudy}
\textbf{ID}     &\textbf{Age}   &\textbf{Gender} & \textbf{Visual Acuity} & \textbf{Impairment} & \textbf{Workshop Part.} \\
\hline\hline

\PoneStudy             & 67            & F      & 0\%           & total blindness & \textbf{$\times$}  \\      
\PtwoStudy             & 53            & F      & 0\%           & total blindness &  \textbf{$\times$} \\    
\PthreeStudy             & 60            & M      & 10\%           & vision becomes gray when in distance &  \\   
\PfourStudy             & 72            & F      & 0\%           & total blindness &  \\  
\PfiveStudy             & 45            & F      & 1.5\%           & only contours visible & \textbf{$\times$} \\  
\PsixStudy             & 62            & M      & 5\%           & total blindness on the left eye, right eye blurry vision & \textbf{$\times$}  \\  
\PsevenStudy             & 65            & M      & 3\%           & blurry vision, black spots in the fovea &    \\  
\PeightStudy             & 29            & M      & 14\%           & tunnel vision &   \\ 
\PnineStudy             & 53            & F      & 2\%           & red. vision in the left eye, only close objects are visible &  \\ 
\PtenStudy             & 68            & F      & 10\%           & blind spots in the fovea  & \\ 
\PelevenStudy             & 23            & F      & 6\%           & blind spots in the fovea & \\ 
\PtwelveStudy             & 71            & M      & 0\%           & total blindness & \\ 
\PthirteenStudy             & 62            & M      & 10\%           & blurry vision & \\ 
\PfourteenStudy             & 76            & F      & 3.5\%           & colors visible but blurry & \textbf{$\times$} \\ 
\PfifteenStudy             & 77            & M      & 12\%           & total blindness in fovea but limited vision in periphery & \\ 
\PsixteenStudy             & 62            & F      &  1\%          &  perception of brightness/darkness & \\

\hline\hline
\end{tabularx} 
\end{table*}

\newpage
\section{Descriptive Data of the User Study}\label{app:descriptive_data_user_study}

\begin{table*}[ht!]
\caption{Table of the descriptive data of the user study}
\Description{Table of the descriptive data of the user study, including the depending variables and the completion time for the task of each condition and scenario.}
% \footnotesize
% \scriptsize
% \small
\begin{tabularx}{\textwidth}{lllrrrrrr}
 \textbf{Variable} & \textbf{System} & \textbf{Scenario} & \textbf{n} & \textbf{Min} & \textbf{Max} & \textbf{Mean} & \textbf{Median} & \textbf{SD} \\ 
  \hline \hline
  \addlinespace[2pt]
        Mental Demand~\cite{hart1988development} & \pathfinder & Urban & 16 & 1.00 & 18.00 & 7.56 & 7.50 & 5.15 \\
                  & \pathfinder & Rural & 16 & 2.00 & 15.00 & 7.94 & 7.50 & 5.34 \\ \cline{2-9}
                  & Auditory & Urban & 16 & 4.00 & 20.00 & 11.25  & 10.00 & 5.08 \\
                  & Auditory & Rural & 16 & 1.00 & 20.00 & 10.94 & 10.00 & 6.29 \\
                  \hline
        Usability (SUS)~\cite{SUS_questionaire} & \pathfinder & Urban & 16 & 20.00 & 80.00 & 64.36 & 67.50 & 16.42 \\
                  & \pathfinder & Rural & 16 & 20.00 & 87.50 & 65.31 & 67.50 & 16.71 \\ \cline{2-9}
                  & Auditory & Urban & 16 & 32.50 & 80.00 & 62.34 & 62.50 & 13.21 \\
                  & Auditory & Rural & 16 & 27.50 & 80.00 & 62.50 & 65.00 & 15.33 \\ 
                  \hline
        Situation Awareness (SART)~\cite{taylor2017situational} & \pathfinder & Urban & 16 & 15.00 & 29.00 & 19.69 & 17.00 & 4.76 \\
                  & \pathfinder & Rural & 16 & 7.00 & 35.00 & 18.50 & 18.00 & 6.27 \\ \cline{2-9}
                  & Auditory & Urban & 16 & 10.00 & 30.00 & 19.06 & 19.00 & 5.30 \\
                  & Auditory & Rural & 16 & 3.00 & 28.00 & 16.75 & 18.00 & 7.06 \\ 
                  \hline
        Perceived Safety~\cite{faas2020longitudinal} & \pathfinder & Urban & 16 & -1.75 & 3.00 & 1.44 & 1.86 & 1.42 \\
                  & \pathfinder & Rural & 16 & -0.50 & 3.00 & 1.50 & 2.00 & 1.21 \\ \cline{2-9}
                  & Auditory & Urban & 16 & -1.50 & 3.00 & 1.25 & 1.25 & 1.47 \\
                  & Auditory & Rural & 16 & -1.50 & 3.00 & 1.61 & 2.13 & 1.50 \\ 
                  \hline \hline
        \textit{Completion Time (in min:sec)} & \pathfinder & Urban & 16 & 0:50 & 4:57 & 1:52 & 1:29 & 1:06 \\
                  & \pathfinder & Rural & 16 & 0:35 & 4:37 & 1:41 & 1:32 & 0:51 \\ \cline{2-9}
                  & Auditory & Urban & 16 & 1:02 & 3:55 & 1:56 & 1:59 & 0:45 \\
                  & Auditory & Rural & 16 & 1:03 & 4:09 & 1:57 & 1:18 & 1:03 \\ 
                  \hline \hline

% haptic gesamt	auditory gesamt
% mean	1:46	1:57
% min	0:35	1:02
% max	4:57	4:09
% median	1:32	1:34
% SD	0.041012456	0.038407651

\label{tab:_descr_stat}
\end{tabularx}

\end{table*}

\end{document}